\documentclass[aps,pra,twocolumn,superscriptaddress,reprint]{revtex4-2}
\usepackage{physics}
\usepackage{graphicx}
\usepackage{amsfonts}
\usepackage[colorlinks=true,citecolor=blue,linkcolor=blue,urlcolor=blue]{hyperref}
\usepackage{bbold} 
\usepackage{xcolor}
\usepackage{xr}

\newcommand{\SF}{S_\mathrm{F}} 

\DeclareMathOperator{\sign}{sign}

\newcommand{\depsEP}{M_\mathrm{s,EP}}
\newcommand{\depsEPmin}{M_\mathrm{s,EP}^{\mathrm{min}}}
\newcommand{\msCPAmin}{M_\mathrm{s,CPA}^{\mathrm{min}}}

\newcommand{\chiz}{\chi_0}
\newcommand{\ms}{M_\mathrm{s}}

\newcommand{\Es}{E} 

\newcommand{\instate}{c^{\mathrm{in}}}
\newcommand{\outstate}{c^{\mathrm{out}}}

\newcommand{\nuz}{{\nu_{0}}}
\newcommand{\nup}{{\nu_{+}}}
\newcommand{\num}{{\nu_{-}}}
\newcommand{\nupm}{{\nu_{\pm}}}
\newcommand{\nump}{{\nu_{\mp}}}

\newcommand{\cin}{c^{\mathrm{in}}}
\newcommand{\cout}{c^{\mathrm{out}}}

\newcommand{\cinoutv}{\vb{c}^{\mathrm{in/out}}}

\newcommand{\cinv}{\vb{c}^{\mathrm{in}}}
\newcommand{\coutv}{\vb{c}^{\mathrm{out}}}

\newcommand{\dinv}{\vb{d}^{\mathrm{in}}}

\newcommand{\doutv}{\vb{d}^{\mathrm{out}}}

\newcommand{\dinoutv}{\vb{d}^{\mathrm{in/out}}}

\newcommand{\indslab}{{\sigma,n}}
\newcommand{\indslabL}{{\mathrm{l},n}}

\newcommand{\tlambda}{\tilde{\lambda}}
\newcommand{\rFt}{\tilde{r}_\mathrm{F}}
\newcommand{\tFt}{\tilde{t}_\mathrm{F}}
\newcommand{\SFt}{\tilde{S}_\mathrm{F}}

\newcommand{\evecL}{\tilde{c}^{\mathrm{in}}} 
\newcommand{\evecR}{c^{\mathrm{in}}} 

\newcommand{\Ln}{L_\mathrm{n}}

\newcommand{\rs}[1]{E^{\mathrm{RS}}_{\mathrm{s},#1}}
\newcommand{\rsomc}[1]{\bar{\omega}_{\mathrm{s},#1}}
\newcommand{\rsom}[1]{\omega_{\mathrm{s},#1}}
\newcommand{\rsgamma}[1]{\gamma_{\mathrm{s},#1}}

\DeclareMathOperator{\wind}{wind}

\newcommand{\pt}{\mathcal{P}\mathcal{T}}

\definecolor{cor_color}{RGB}{233,113,50}

\begin{document}
\title{Exceptional Points, Lasing, and Coherent Perfect Absorption \\ in Floquet Scattering Systems}
\author{David~Globosits}
\thanks{These authors contributed equally: David Globosits, Puneet Garg}
\affiliation{Institute for Theoretical Physics, Vienna University of Technology (TU Wien), 1040 Vienna, Austria}

\author{Puneet~Garg}
\thanks{These authors contributed equally: David Globosits, Puneet Garg}
\affiliation{Institute of Theoretical Solid State Physics, Karlsruhe Institute of Technology, 76131 Karlsruhe, Germany}
\author{Jakob~Hüpfl}
\affiliation{Institute for Theoretical Physics, Vienna University of Technology (TU Wien), 1040 Vienna, Austria}
\author{Adrià~Canós~Valero}
\affiliation{Institute of Physics, University of Graz, and NAWI Graz, 8010 Graz, Austria}
\affiliation{Riga Technical University, Institute of Telecommunications, 1048 Riga, Latvia}
\author{Thomas Weiss}
\affiliation{Institute of Physics, University of Graz, and NAWI Graz, 8010 Graz, Austria}
\author{Carsten~Rockstuhl}
\affiliation{Institute of Theoretical Solid State Physics, Karlsruhe Institute of Technology, 76131 Karlsruhe, Germany}
\affiliation{Institute of Nanotechnology, Karlsruhe Institute of Technology, 76131 Karlsruhe, Germany}
\author{Stefan~Rotter}
\affiliation{Institute for Theoretical Physics, Vienna University of Technology (TU Wien), 1040 Vienna, Austria}

\begin{abstract} 
Periodically time-varying media, known as photonic time crystals (PTCs), provide a promising platform for observing unconventional wave phenomena. We analyze the scattering of electromagnetic waves from spatially finite PTCs using the multispectral Floquet scattering matrix, which naturally incorporates the frequency-mixing processes intrinsic to such systems. For dispersionless, real, and time-periodic permittivities, this matrix is pseudounitary. Here, we demonstrate that this property leads to multiple symmetry-breaking transitions: for increasing driving strength, scattering matrix eigenvalues lying on the unit circle (unbroken symmetry regime) meet at exceptional points (EPs), where they break up into inverse complex conjugate pairs (broken symmetry regime). We identify the symmetry operator associated with these transitions and show that, in time-symmetric systems, it corresponds to the time-reversal operator. Remarkably, at the parametric resonance condition, one eigenvalue vanishes while its partner diverges, signifying simultaneous coherent perfect absorption (CPA) and lasing. Since our approach relies solely on the Floquet scattering matrix, it is not restricted to a specific geometry but instead applies to any periodically time-varying scattering system. To illustrate this universality, we apply our method to a variety of periodically time-modulated structures, including slabs, spheres, and metasurfaces. In particular, we show that using quasi-bound states in the continuum resonances sustained by a metasurface, the CPA and lasing conditions can be attained for a minimal modulation strength of the permittivity. Our results pave the way for engineering time-modulated photonic systems with tailored scattering properties, opening new avenues for dynamic control of light in next-generation optical devices.
\end{abstract}

\maketitle

\section{Introduction}
Electromagnetic waves inside dielectric media with a time-varying permittivity have recently been found to produce a host of unconventional wave phenomena~\cite{Galiffi22}, such as magnetic-free non-reciprocal energy transfer~\cite{Sounas,Li18,Li182,Li19}, enhanced photon pair generation~\cite{Prain17}, and exotic topological phases~\cite{Ni25}, among others. Recently, time-varying media were also employed for amplifying or attenuating~\cite{hayran2024beyond,suwunnarat2019dynamically} incoming radiation. Especially useful in this regard are media with periodically time-varying properties, so-called Floquet media or photonic time crystals (PTCs)~\cite{Lustig232,Asgari:24}. Several theoretical studies, in which the amplification of waves inside bulk PTCs has been discussed~\cite{Lyubarov,Khurgin25}, were recently accompanied by innovative experiments realizing extreme amplification of waves inside time-varying media~\cite{Wang23}. Complementary demonstrations of coherent absorption in such media have likewise been reported~\cite{Galiffi24}.

Capturing these developments calls for a unified theoretical framework that accounts for the distinctive features of time-varying media, including frequency conversion, amplification, and attenuation. In practice, realizations of PTCs are always spatially finite, which requires additional considerations beyond the bulk description. A comprehensive framework must therefore also incorporate the role of finite size and geometry, which govern processes such as boundary-induced scattering and resonance effects. These effects can profoundly modify the response of a finite PTC compared to its bulk counterpart~\cite{Stefanou,Wang25,Valero25,garg2025photonic,Verde}. Furthermore, for finite Floquet media the crucial question arises, which specific far-field input state couples most efficiently to the time-varying medium to achieve extreme wave amplification or attenuation. Answering this question provides a direct connection between the physics of finite PTCs and established wavefront-shaping approaches~\cite{Cao22}.

In this work, we provide such a unifying description using the Floquet scattering matrix~\cite{Reichl99}. This matrix comprehensively characterizes the periodically time-varying scattering process, expressing how incoming light fields are scattered into outgoing light fields incorporating reflections at the system's  boundaries and the geometry-specific resonances these boundaries give rise to~\cite{Zurita,Martinez16,Martinez18,Pantazopoulos,buddhiraju2021arbitrary,Fan}. This makes the Floquet scattering matrix advantageous over conventionally used band structure descriptions of bulk PTCs~\cite{Asgari:24}. Even a generalized band structure analysis that characterizes the resonant states of periodically driven, open scattering systems~\cite{Wang25,Valero25} does not provide direct access to the output field generated by a given input field in a finite structure (see Sec.~\ref{app:bandstructures} in the Supplementary Information).

We also emphasize that in time-varying scattering setups energy and therefore frequency are not conserved quantities. The Floquet scattering matrix correctly captures this frequency mixing induced by time modulation. Particularly important are those processes in which positive and negative frequencies are coupled with each other, which is typically achieved for systems with a strong and fast modulation~\cite{Serra}. As was recently demonstrated~\cite{Globosits}, these processes are responsible for a special algebraic property of the Floquet scattering matrix: If the system is free of losses and dispersion, the Floquet scattering matrix is a pseudounitary matrix, expressing that the wave action, also known as the number of pseudophotons~\cite{Pendry23}, is a conserved quantity.

Here, we demonstrate that the pseudounitarity of the Floquet scattering matrix has significant implications for the physics of periodically time-varying media. In particular, we find that the distribution of the eigenvalues and eigenstates reveals a close connection between non-Hermitian, $\pt$-symmetric scattering systems and Floquet scattering systems. We show that for weak temporal modulations, all eigenvalues of the corresponding scattering matrix are unimodular. Quite remarkably, by increasing the modulation strength, the system can undergo multiple symmetry-breaking transition at exceptional points (EPs), where two eigenvalues coincide and the associated eigenvectors become parallel. We show that for time-symmetric modulations of the permittivity, the time-reversal symmetry of the eigenstates is broken. By increasing the modulation strength even further, eigenvalues leave the unit circle pairwise.

At the parametric resonance condition, one eigenvalue eventually can become zero while its partner eigenvalue simultaneously diverges to infinity. Such an operational condition marks the point where the system can act as both a coherent perfect absorber (CPA) and a laser. The symmetry-breaking transition at an EP and the fact that the system can operate as a CPA-laser are strongly reminiscent of what has been found for $\pt$-symmetric scattering systems~\cite{Longhi,Chong}.

Recently, several studies have revealed important insights into the role of symmetry breaking at EPs and their connection to extreme energy transfer in unbounded (bulk) time-modulated media~\cite{Kazemi,Koutserimpas20,Li,zhang2025}. However, to accurately describe phenomena such as CPA and lasing, a framework is required that accounts for the finite size of any practical realization of a time-varying medium. In particular, the presence of spatial boundaries leads to scattering and interference effects that are absent in unbounded systems. As our framework is based on the Floquet scattering matrix, it not only captures the unique features of periodically time-varying systems but also naturally incorporates the effects of a finite system size independent of the complexity of the setup. With our approach being very general, it applies to a wide range of time-modulated scattering systems, given that a corresponding Floquet scattering matrix can capture their response. It incorporates an arbitrary number of Floquet channels and holds true for any frequency of the incident light field.

Furthermore, in contrast to bulk media, in finite Floquet systems, specific far-field input states experience an extreme amplification or attenuation while other states conserve the number of photons. Our formalism explicitly accounts for how far-field input wave field couple to the time-varying medium, which allows us to identify these special wave fields as the eigenvectors of the Floquet scattering matrix. Notably, this includes the wave fields that get emitted by the Floquet laser and that get perfectly absorbed by the Floquet CPA. Since the Floquet scattering matrix incorporates the scattering behavior of multiple frequencies, the eigenstates and, therefore, also the lasing and CPA states are multi-spectral and, thus, in general, pulsed light fields.

Our manuscript is structured as follows: We first discuss our framework based on a finite slab with oscillating permittivity. Then, we apply it to spatially localized scatterers and finally to a periodic arrangement of scatterers in two dimensions, i.e., a metasurface. With that, we show that all the considered Floquet scattering systems share the same underlying physics.

\section{Results}
\subsection{The Floquet Scattering Matrix $\SF$}
\begin{figure*}
    \includegraphics[width=.95\textwidth]{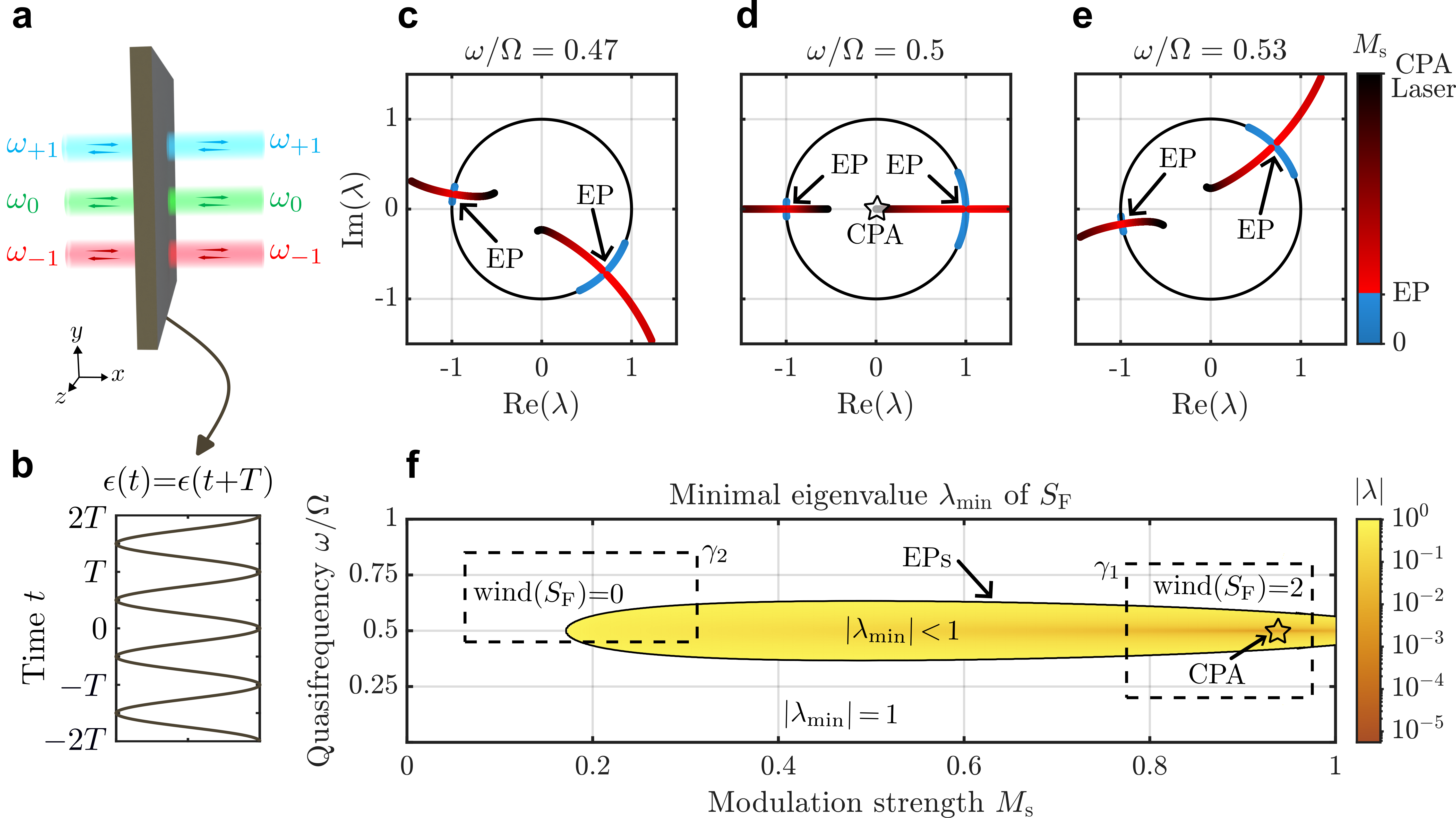}
    \caption{\label{fig:slab_Fig1} Behavior of the eigenvalues $\lambda$ of the Floquet scattering matrix $\SF$ for a time-varying slab (see text for parameter values). \textbf{a},~We consider light scattering off a slab with a time-periodic permittivity with period $T=2\pi/\Omega$. The frequency of the incident light field can change during scattering such that the output light consists of several frequency components $\omega+n\Omega$. \textbf{b},~We assume a time-harmonic permittivity of the slab. \textbf{c-e},~We track four eigenvalues of $\SF$ as a function of the modulation strength $\ms$ for different choices of the quasifrequency $\omega$. In all cases, for small $\ms$ all eigenvalues are unimodular and thus lie on the unit circle (blue). At a critical modulation strength, EPs are formed, where two respective eigenvalues coincide. By further increasing $\ms$, these two eigenvalues leave the unit circle in a pairwise fashion (red). The colorbar to the right of panel \textbf{e} applies to all panels \textbf{c-e}, schematically indicating the respective EPs by a blue-to-red transition (the actual $\ms$ value at which the EPs occur differs between panels). \textbf{d},~A special situation occurs when the quasifrequency of the wave field obeys the parametric resonance condition, $\omega/\Omega=0.5$. Then, one eigenvalue vanishes (represented by a star symbol) while another eigenvalue diverges (not shown), at which point the Floquet system acts as both a CPA and a laser. \textbf{f},~The absolute value of the minimal eigenvalue $\lambda_\mathrm{min}$ of $\SF$ is shown here as a function of the modulation strength $\ms$ and the quasifrequency $\omega$. For small $\ms$ (weak modulation) or if the quasifrequency is far detuned from the parametric resonance condition, the minimal eigenvalues, and therefore all eigenvalues, are unimodular (white). This unbroken regime is separated by an exceptional line (black) from the broken regime with $\abs{\lambda_\mathrm{min}}<1$ (yellow). When calculating the winding number along the path $\gamma_1$, we find $\wind(\SF) = 2$, verifying that this path encloses a CPA-lasing point (star symbol). For reference, we also calculate the winding number for the path $\gamma_2$, which does not enclose a CPA-lasing point, and find $\wind(\SF)=0$.}
\end{figure*}

For the sake of simplicity, we first consider linearly polarized light (in $z$-direction) at normal incidence scattering off a dielectric slab that is infinite in $y$- and $z$-directions and has finite thickness $L$ in $x$-direction [see Fig.~\ref{fig:slab_Fig1}a]. Furthermore, we assume that the slab's material responds instantaneously to the external field. The scattering system can then be described with a real, time-periodic dielectric function $\epsilon(\vb{r},t)=\epsilon(\vb{r},t+T)$, where $T$ is the oscillation period and $\Omega=2\pi/T$ is the associated angular frequency. Specifically, we choose a time-harmonic modulation of the permittivity [see Fig.~\ref{fig:slab_Fig1}b]. We further assume that the slab is surrounded by free space. This scattering system constitutes an effective one-dimensional Floquet scattering problem governed by the scalar wave equation for the complex-valued electric field $\Es(x,t)$ as
\begin{equation}
    \partial_x^2 \Es(x,t)-\frac{1}{c^2 }\partial_t^2 \left[ \epsilon(x,t) \Es(x,t) \right]=0 \label{eq:scalar_waveeq_E}
\end{equation}
where $c$ is the speed of light in free space, and the relative permittivity is given by
\begin{equation}
    \epsilon(x,t)=1+\chiz [1 - \ms \cos(\Omega t)]\Theta(L/2-\abs{x}) \label{eq:epsilon_1d}
\end{equation}
Here, $\Theta$ is the Heaviside step function, $L$ is the thickness of the scatterer, $1+\chiz$ is the static permittivity of the slab, and $\ms \geq 0$ is the modulation strength (modulation amplitude). If not stated otherwise, we assume for the parameters of the slab $\chiz=4$, $\Omega=2\pi \times 151$~THz, and $L=1581$~nm.

The wave fields in the asymptotic regions left ($x~<~-L/2$, $\sigma=\mathrm{l}$) and right ($x>L/2$, $\sigma=\mathrm{r}$) of the slab are assumed to be superpositions of plane waves with frequencies $\omega_n=\omega+n\Omega$ with $n \in \mathbb{Z}$. Here, $\omega$ is the quasifrequency (Floquet frequency) which we choose to lie in the first temporal Brillouin zone $0\leq \omega < \Omega$. The incoming and outgoing wave fields expressed in a photon-flux normalized basis read~\cite{Globosits}
\begin{subequations}
    \begin{align}
    E^{\mathrm{in}}(x,t) &= \sum_\indslab \sqrt{\hbar \mu_0 c \abs{\omega_n}} \cin_\indslab e^{i k_\indslab (x-L/2)} e^{- i\omega_n t}  \\
    E^{\mathrm{out}}(x,t) &= \sum_\indslab \sqrt{\hbar \mu_0 c \abs{\omega_n}} \cout_\indslab e^{-i k_\indslab (x-L/2)} e^{- i\omega_n t} 
    \label{eq:asym_Efield_slab}
    \end{align}
\end{subequations}
In the asymptotic regions, a linear dispersion relation holds such that $c k_{\mathrm{l},n}=-c k_{\mathrm{r},n}=\omega_n$. Furthermore, we introduce the amplitudes $\cin_\indslab$ and $\cout_\indslab$ for incoming and outgoing plane waves at the discrete frequencies $\omega_n=\omega+n\Omega$. The Floquet scattering matrix connects the incoming with the outgoing wave fields as
\begin{equation}
    \ket*{\outstate}=\SF \ket*{\instate} 
\end{equation}
Here, we arrange the amplitudes into vectors as $\ket*{\instate}=(\dots,c^{\mathrm{in}}_{\mathrm{l},-1},c^{\mathrm{in}}_{\mathrm{l},0},\dots,c^{\mathrm{in}}_{\mathrm{r},-1},c^{\mathrm{in}}_{\mathrm{r},0},\dots)^\mathrm{T}$ and analogously for the outgoing state $\ket*{\outstate}$. We refer the reader to~\cite{Globosits,Zurita} for details on how to numerically obtain the respective Floquet scattering matrix for the system at hand. For computational reasons, we truncate the system to $n \in [-N,N-1]$ Floquet channels. Specifically, for the slab system, we choose $N=8$. When the Floquet scattering matrix is expressed in a photon-flux normalized basis, it obeys a pseudounitary relation of the form~\cite{Globosits}
\begin{equation}
    \SF^{\dagger}V\SF=V  \label{eq:pseudounitarity}
\end{equation}
where
\begin{equation}
    V = \mqty(-\mathbb{1}  & \mathbb{0} & \mathbb{0} & \mathbb{0} \\
          \mathbb{0} & \mathbb{1} & \mathbb{0} & \mathbb{0} \\
          \mathbb{0} & \mathbb{0} & -\mathbb{1} & \mathbb{0} \\
          \mathbb{0} & \mathbb{0} & \mathbb{0} & \mathbb{1} )  \label{eq:V_def}
\end{equation}
The appearing matrices ($\pm \mathbb{1}$ and $\mathbb{0}$) are of appropriate size to ensure that the matrix $V$ assigns a minus sign to negative frequency channels ($n<0$) via the matrices $-\mathbb{1}$ and a plus sign to positive frequency channels ($n \geq 0$) via the matrices $\mathbb{1}$ on each side of the slab. Specifically, the upper left (lower right) matrix quadrant corresponds to the left (right) port.

\subsection{Exceptional Points and the CPA-Lasing Condition}
In this work, we demonstrate that the pseudounitary property of $\SF$ has profound consequences for the physics of Floquet scattering systems. In particular, we show that effects like EPs, lasing, and coherent perfect absorption, which are impossible to obtain with wave fields associated with real frequencies in energy-conserving, time-invariant systems, can emerge in Floquet scattering systems. Previously, these effects were observed in non-Hermitian systems that break energy conservation with static gain and loss elements, rendering the system non-unitary, or using complex frequency input waves~\cite{Baranov,Valero24}. Here, we reveal that such phenomena also emerge naturally in spatially finite time-varying systems.

In energy-conserving systems, the scattering matrix is unitary when expressed in an energy-flux normalized basis, and thus all eigenvalues $\lambda$ are unimodular $\abs{\lambda_n}=1$. This property is altered for pseudounitary Floquet scattering matrices, which we can understand by the following observation~\cite{Mostafazadeh}: Let $\ket*{\instate}$ be an eigenstate of $\SF$ with corresponding eigenvalue $\lambda$. Then, from Eq.~\eqref{eq:pseudounitarity}, we see that $V\ket*{\instate}$ is an eigenstate of $\SF^\dagger$ with eigenvalue $1/\lambda$ as
\begin{equation}
        \SF^{\dagger}V\ket*{\instate}=V\SF^{-1}\ket*{\instate} = \frac{1}{\lambda}V\ket*{\instate} 
\end{equation}
Since the eigenvalues of $\SF^{\dagger}$ are the complex conjugates of the eigenvalues of $\SF$, we arrive at the result that both $\lambda$ and $1/\lambda^*$ are eigenvalues of $\SF$. Importantly, we conclude that two different regimes are possible: First, eigenvalues may be unimodular, i.e., $\abs{\lambda}=1$, such that $\lambda=1/\lambda^*$. When such a condition is fulfilled, the corresponding eigenstate is in the so-called unbroken regime. By contrast, in the broken regime, eigenvalues are not unimodular ($\abs{\lambda} \neq 1$), but instead come in inverse complex conjugate pairs $(\lambda, 1/\lambda^*)$. These two regimes are separated by a spontaneous symmetry-breaking transition occurring at an exceptional point where two eigenvalues coalesce, and the respective eigenstates of $\SF$ become parallel (we numerically checked that the eigenstates indeed become parallel at an EP for all scattering setups~\cite{Schomerus}).

Notably, this algebraic property of the eigenvalues of a scattering matrix can also be found for $\pt$-symmetric scattering systems~\cite{Ozdemir}. There, static gain and loss elements are arranged in a spatially symmetric configuration, such that the scattering landscape is invariant under the combined action of the parity operator and the time-reversal operator. The eigenvalues of the corresponding scattering matrix are also either unimodular (unbroken $\pt$-symmetry regime) or come in inverse complex conjugate pairs (broken $\pt$-symmetry regime). Similar to the Floquet case, both regimes are separated by an exceptional point indicating the $\pt$-symmetry-breaking transition. Remarkably, a special situation may also occur in which one eigenvalue vanishes while at the same time one eigenvalue diverges. This marks the situation at which the system can simultaneously act as a CPA (vanishing eigenvalue) and as a laser (diverging eigenvalue)~\cite{Longhi,Chong,AmbichlPRX,Wong}.

In the following, we show that similar unconventional wave phenomena appear in Floquet scattering systems, which can be understood based on the eigenvalues of the associated Floquet scattering matrix. To demonstrate this, we track the behavior of the eigenvalues of $\SF$ as a function of the modulation strength $\ms$. In Fig.~\ref{fig:slab_Fig1}c-e, we depict those four eigenvalues out of all the eigenvalues of $\SF$ that undergo a symmetry-breaking transition with increasing $\ms$ (blue-red colorscale) for the chosen set of parameters for three different choices of the quasifrequency $\omega$. In all three cases, we observe that if the modulation is weak, the eigenvalues lie on the unit circle (represented in blue). In fact, all the eigenvalues of $\SF$ are unimodular in this regime (not shown). 

With increasing $\ms$, the eigenvalues start to shift while remaining on the unit circle (blue colorscale). Specifically, we observe in Fig.~\ref{fig:slab_Fig1}c-e that two eigenvalues with positive real parts and two other eigenvalues with negative real parts approach each other, respectively. Once a critical modulation strength is reached, the nearby eigenvalues coalesce, marking the formation of two EPs. We notice that these two EPs form at the same modulation strength $\ms$. By increasing $\ms$ even further, the four depicted eigenvalues leave the unit circle and form two pairs, each of which contains partners that are inverse complex conjugate to each other, $(\lambda,1/\lambda^*)$ (red colorscale). This marks the situation where the associated eigenstates enter the broken regime.

A remarkable situation occurs when the quasifrequency meets the parametric resonance condition, $\omega/\Omega~=~1/2$, as depicted in Fig.~\ref{fig:slab_Fig1}d. There, one eigenvalue in the broken regime vanishes while its partner eigenvalue diverges. At this particular driving strength, the slab behaves as a CPA ($\lambda=0$) and simultaneously as a laser ($\lambda=\infty$). 

For comparison, we also show the eigenvalue structure for the off-resonant case $\omega/\Omega<1/2$ [see Fig.~\ref{fig:slab_Fig1}c] and $\omega/\Omega>1/2$ [see Fig.~\ref{fig:slab_Fig1}e], where an eigenvalue becomes small (large) but does not reach zero (infinity), which highlights the importance of the parametric resonance condition not only for bulk but also for spatially finite PTCs to reach the condition for CPA or lasing, respectively. We note that all the other eigenvalues that are not depicted in Fig.~\ref{fig:slab_Fig1}c-e stay on the unit circle throughout the whole interval of modulation strength $\ms$ for the chosen set of parameters. In general, this means that some eigenstates of $\SF$ can be in the broken phase while other eigenstates for the same set of parameters are in the unbroken phase. It would be interesting to explore special situations, such as when multiple EPs form at the same position in parameter space using, for instance, tools from Krein stability analysis~\cite{Flynn}.

To provide further insights on the influence of the quasifrequency $\omega$ and the modulation strength $\ms$ for the formation of EPs and the CPA-lasing threshold, we plot the absolute value of the smallest eigenvalue of $\SF$ in Fig.~\ref{fig:slab_Fig1}f. We observe that for a weak driving strength and also for light fields that are far detuned from the parametric resonance condition $\omega/\Omega=1/2$, the smallest eigenvalue (and therefore all eigenvalues) are unimodular (white color). On the contrary, if the quasifrequency is close to the parametric resonance condition, by increasing $\ms$, the eigenvalues can reach an exceptional point (black line). Figure~\ref{fig:slab_Fig1}f suggests that a finite modulation strength is necessary to observe an EP. However, by properly adjusting the thickness of the slab, an EP can form for arbitrarily small modulation strengths -- a result that may be very relevant for the experimental observation of EPs in spatially finite Floquet media. More specifically, using a two-band model approximation under the assumption of a weak modulation strength and setting $\omega/\Omega=1/2$, we derive an approximate expression for the minimal modulation strength necessary to observe an exceptional point $\depsEPmin$. We find that
\begin{equation}
\depsEPmin \propto \sqrt{1-\cos(\Ln)}  \label{eq:deps_EP}
\end{equation}
where we introduced the normalized thickness $\Ln=L\Omega\sqrt{1+\chiz}/c$. Interestingly, Eq.~\eqref{eq:deps_EP} tells us that $\depsEPmin$ vanishes for $\Ln= 2\pi m$ with $m \in \mathbb{N}$, which exactly corresponds to the resonances of the static slab. Hence, the resonant response is crucial for reducing the modulation strength required to reach the EP. We verify the above observation using the numerical data of our full-scale simulation, which takes more than two frequency components into account (we use $n \in [-8,7]$) and does not rely on the assumption of a weak driving strength~\cite{Globosits}. Especially around $\Ln= 2\pi m$, we find excellent agreement between the results derived from the linearized two-band model approximation and the full-scale simulation as the assumption of weak modulation is well satisfied in these cases. For details on the derivation of Eq.~\eqref{eq:deps_EP} including the full expression of $\depsEPmin$ and a comparison of the approximate result to the non-perturbative data, we refer the reader to Subsec.~\ref{app:tbm_EP}. 

Figure~\ref{fig:slab_Fig1}(f) further shows that by increasing $\ms$ and thus going beyond the EP, the smallest eigenvalue enters the broken regime obeying $\abs{\lambda_\mathrm{min}} < 1 $ (yellow colorscale). In the specific case when $\omega/\Omega=1/2$, the smallest eigenvalue becomes zero (marked by the star symbol), thus reaching the CPA condition. By increasing $\ms$ even further, the magnitude of the minimal eigenvalue increases again. We note that in the regime beyond the CPA-lasing condition, non-linear effects can be expected to arise, which are not described by the linear model we assume here~\cite{Esterhazy,Tureci}. In Subsec.~\ref{app:tbm_CPA}, we provide an analysis of the minimal driving strength necessary for the system to operate as a CPA or as a laser, respectively.

\begin{figure}
    \includegraphics[width=\columnwidth]{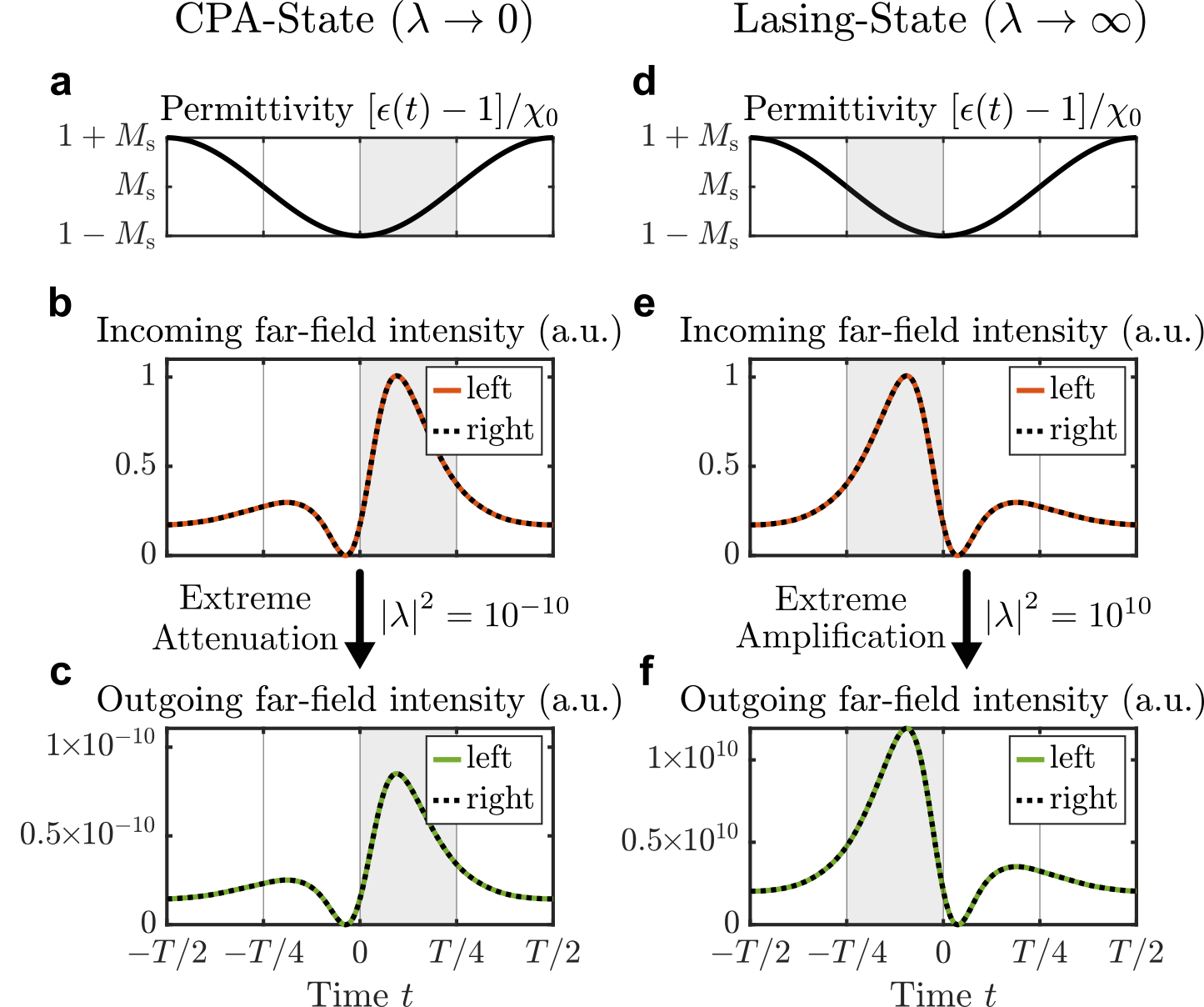}
    \caption{\label{fig:slab_Intensity_borders} Intensities of the incoming and the outgoing CPA and lasing states at the borders of the slab (see text for parameter values). \textbf{a}~and~\textbf{d},~Temporal variation of the periodic permittivity function of the slab. \textbf{b},~Intensity of the incoming part of the CPA light field at the left border at $x=-L/2$ (red) and at the right border at $x=L/2$ (black) of the slab. The light field approaches the slab from the left and right with the same temporal intensity profile (the two lines overlap). Destructive interference of these pulses eliminates spatial reflections. The wave field builds up a large intensity maximum at times, when the permittivity is rising $0\leq t \leq T/4$ (gray shaded region). In this way, all the energy of the light field gets perfectly absorbed by the time-varying medium. \textbf{c},~The corresponding output intensity at the borders of the slab is reduced by a factor $10^{-10}$. \textbf{e},~Same as \textbf{b} but for the lasing state. This light field builds up large intensity maxima at times, when the permittivity is lowered $-T/4 \leq t \leq 0$ (gray shaded region), receiving energy from the time-modulated slab. \textbf{f},~This leads to a huge amplification of the outgoing intensity by a factor of $10^{10}$. Furthermore, we observe the time-reversal symmetry of the CPA and lasing states: The incoming CPA state is the time-reversed of the outgoing lasing state.}
\end{figure}

The eigenvector corresponding to the vanishing eigenvalue is a light field that, when injected into the scattering system, gets perfectly absorbed inside the time-varying medium such that no outgoing (reflected or transmitted) wave is produced. In Fig.~\ref{fig:slab_Intensity_borders}a-c, we demonstrate how the CPA-state manages to be perfectly absorbed by plotting the intensities of the electric field at the left ($x=-L/2$) and right ($x=L/2$) interfaces of the slab. Since no evanescent modes are excited in this one-dimensional slab system, the wave field at the border of the slab can already be considered as the far-field. 

To be perfectly absorbed, the input field depicted in Fig.~\ref{fig:slab_Intensity_borders}b is optimally shaped in both its spatial and temporal degrees of freedom. The spatial degrees of freedom are adjusted so that all reflections off the spatial interfaces of the slab are eliminated through destructive interference. Such an operation is accomplished by simultaneously approaching the scatterer from left and right, which we observe in Fig.~\ref{fig:slab_Intensity_borders}b by noting that the intensity distributions at the left and right interfaces are identical. On the other hand, the temporal degrees of freedom are fine-tuned in such a way that only minimal intensity is built up at the interfaces during the first half of the period $-T/2 \leq t \leq 0$. However, precisely during the duration when the permittivity starts to rise $0 < t \leq T/4$ [gray region, see also Fig.~\ref{fig:slab_Intensity_borders}a], this light field hits the scatterer and enters the time-varying slab corresponding to the intensity maximum during this time frame in Fig.~\ref{fig:slab_Intensity_borders}b. A large intensity build-up occurs inside the Floquet slab. In this way, the energy of the light field gets completely absorbed by the time-varying medium, and nearly no outgoing wave is produced, as shown in Fig.~\ref{fig:slab_Intensity_borders}c. 

Conversely, we plot the incoming temporal intensity distribution at the interfaces of the slab for the lasing state in Fig.~\ref{fig:slab_Intensity_borders}e. This state exhibits a strong intensity maximum at times when the permittivity is lowered during the interval $-T/4 \leq t \leq 0$ [gray region, see also Fig.~\ref{fig:slab_Intensity_borders}d]. In this way, the light field receives energy from the time-varying medium and thus gets amplified [see~Fig.~\ref{fig:slab_Intensity_borders}f]. Therefore, by approaching the lasing condition, the eigenstate to this diverging eigenvalue will be the dominant contribution at the output for arbitrary input states (excluding the one case of having the corresponding CPA state as the input state). We note that to avoid numerical instabilities in the close proximity of divergences, we plot an eigenstate corresponding to a small but finite eigenvalue $\abs{\lambda}\approx 10^{-5}$ in Fig.~\ref{fig:slab_Intensity_borders}b-c. Correspondingly, the lasing state depicted in Fig.~\ref{fig:slab_Intensity_borders}(e)-(f) corresponds to a large but finite eigenvalue of $\abs{\lambda}\approx 10^{5}$. Since we plot the intensity of the wave field, which is the square of the electric field, the scaling factor in Fig.~\ref{fig:slab_Intensity_borders}c is about $10^{-10}$ and in Fig.~\ref{fig:slab_Intensity_borders}f about $10^{10}$.

To prove unambiguously that in the vicinity of this small but finite eigenvalue there truly exists a vanishing eigenvalue together with a diverging eigenvalue, we calculate the associated winding number $\wind(\SF)$ along two loops $\gamma_{1,2}$ parametrized by $0\leq \phi < 2\pi$ in parameter space [see Fig.~\ref{fig:slab_Fig1}f]. The winding number vanishes if the loop $\gamma$ does not enclose a zero or diverging eigenvalue. If, however, the loop $\gamma$ encloses a CPA-lasing point, we expect a winding number of $\pm2$~\cite{Sakotic}. While originally introduced for single-frequency scattering matrices~\cite{Guo}, we here extend the concept of the winding number to multispectral Floquet scattering matrices. Using the pseudounitarity of $\SF$, the winding number takes the form
\begin{equation}
    \wind(\SF) = \frac{1}{2\pi i} \int_{0}^{2\pi} \dd{\phi} \Tr{V\SF^{\dagger} V \dv{\SF}{\phi}} 
\end{equation}
We indeed numerically confirm that $\wind(\SF) \approx 2 + 4.4 \times 10^{-5}$ along $\gamma_1$, verifying the existence of both a zero and a pole of $\SF$. For comparison, we also calculate the winding number for a loop $\gamma_2$ that does not encircle a zero and pole, where we find $\wind(\SF) \approx 0 + 1.9 \times 10^{-5}$, as expected.

In this subsection, we showed how the eigenvalues of the pseudounitary Floquet scattering matrix behave as a function of the driving strength $\ms$ and the quasifrequency $\omega$. We revealed the appearance of EPs and identified the special cases of CPA and lasing. Notably, since the Floquet scattering matrix is pseudounitary for any real, dispersionless, and time-periodic permittivity function, our framework holds true not only for time-harmonic driving protocols but for arbitrary periodic modulations. In analogy to static, $\pt$-symmetric systems, where an EP marks the breaking of the $\pt$ symmetry of the corresponding eigenstates, we also expect in the time-varying Floquet case a corresponding symmetry-breaking transition to occur at an EP. In the following, we address the question of which symmetry can be spontaneously broken in a Floquet scattering system and which consequences this entails.

\subsection{The Symmetry-breaking Transition \label{sec:symmetry_breaking_transition}}

Spontaneous symmetry-breaking, in our case, means that the eigenstates of $\SF$ in the unbroken regime individually possess a symmetry. In contrast, in the broken regime, this symmetry maps one eigenstate onto its partner and vice versa. As we show below, for a time-symmetric driving protocol $\epsilon(t)=\epsilon(-t)$, the associated symmetry that is broken at an EP is the time-reversal symmetry. 

To make this argument more transparent, we introduce the following notation from Ref.~\onlinecite{Mostafazadeh03} to distinguish between states in the unbroken and the broken scattering regime: We label an eigenstate corresponding to a unimodular eigenvalue as $\ket*{\instate_\nuz}$ with $\abs{\lambda_\nuz}=1$ and the two eigenstates corresponding to a distinct pair of non-unimodular eigenvalues as $\ket*{\instate_\nupm}$ with $\abs{\lambda_\nup}>1$ and $\abs{\lambda_\num}<1$, respectively. Furthermore, we assume that the eigenstates are non-degenerate. For every diagonalizable pseudounitary matrix, we can construct an anti-linear symmetry operator $X$ that has the above-described property~\cite{Mostafazadeh03}
\begin{equation}
    X \ket*{\instate_{\nu}} =
    \begin{cases}
        \ket*{\instate_{\nuz}}\, , \quad & \nu=\nuz  \\
        \ket*{\instate_{\nupm}}\, , \quad & \nu=\nump 
    \end{cases} \label{eq:X_definition}
\end{equation}
Notably, this anti-linear operator $X$ can be constructed for any diagonalizable pseudounitary scattering matrix $\SF$ and thus even for time-periodic driving schemes $\epsilon(t)$ that do not possess any additional temporal symmetry beyond periodicity. We refer the reader to Subsec.~\ref{app:symmetry_operators} for the treatment of this general case, including also a discussion on how to deal with degeneracies of $\SF$. Furthermore, our formalism is fully general as it holds for an arbitrary choice of the quasifrequency $\omega$ and for an arbitrary number of Floquet channels considered in the corresponding wave field. In this way, we provide a complete and comprehensive description of spontaneous symmetry breaking in Floquet scattering systems, including CPA and lasing. The results of Ref.~\onlinecite{Koutserimpas183}, obtained for the restricted case of two Floquet channels at a fixed quasifrequency $\omega/\Omega=0.5$, thus appear as one particular instance of our general framework.

The operator $X$ has a particularly straightforward interpretation for time-reversal symmetric driving protocols $\epsilon(t)=\epsilon(-t)$, which we will assume in the following. In this case, the symmetry represented by the operator $X$ is the time-reversal symmetry, and the associated time-reversal operator is the complex conjugation operator $X=K$, such that when acting on an arbitrary input state $\ket*{\cin}$ we have $K\ket*{\cin}=\ket*{\cin}^*$. To explicitly show that $K$ is the correct symmetry operator, we first note that for time-symmetric driving protocols the Floquet scattering matrix can be chosen to satisfy a generalized reciprocity relation $\SF = V \SF^\mathrm{T} V $ or equivalently $\SF^* = V \SF^\dagger V$ (see Subsec.~\ref{sec:reciprocity} for a derivation of this result). Together with the pseudounitarity condition [Eq.~\eqref{eq:pseudounitarity}], we thus have
\begin{equation}\label{eq: Smat_cc_sym}
    \SF^{-1} = \SF^* \equiv K \SF K 
\end{equation}
Using the above equation, we can see that the operator $K$ has the desired properties of a symmetry operator [Eq.~\eqref{eq:X_definition}] by the following argument: If $\ket*{\instate_{\nu}}$ is an eigenvector of $\SF$ to an eigenvalue $\lambda_{\nu}$ then $K \ket*{\instate_{\nu}}$ is an eigenvector to the eigenvalue $1/\lambda_{\nu}^*$ as
\begin{equation}\label{eq: time-reversal}
\begin{split}
    \SF K \ket*{\instate_{\nu}} &= K \left( \SF^{-1} \ket*{\instate_{\nu}} \right) = \frac{1}{\lambda_{\nu}^*} K \ket*{\instate_{\nu}} 
\end{split}    
\end{equation}
Note that in the unbroken regime where $\abs{\lambda_{\nuz}}=1$, we have $1/\lambda_{\nuz}^* = \lambda_{\nuz}$. The above shows that states in the unbroken regime are time-reversal symmetric individually. On the contrary, an eigenstate $\ket*{\instate_{\nup}}$ in the broken regime is the time-reversed of the corresponding partner state $\ket*{\instate_{\num}}$. In particular, this means that the CPA input state ($\lambda=0$) in Fig.~\ref{fig:slab_Intensity_borders}b is the time-reversed of the output lasing state ($\lambda=\infty$) depicted in Fig.~\ref{fig:slab_Intensity_borders}f, respectively.

It may seem counterintuitive that, on the one hand, a periodically time-varying medium can reduce or enhance the intensity of the outgoing light field, including the extreme cases of CPA and lasing. On the other hand, however, the number of pseudophotons flowing into the scattering system equals the number of pseudophotons flowing out of the system for an arbitrary input state $\ket*{\instate}$~\cite{Pendry23} as expressed by the pseudounitarity of the Floquet scattering matrix [see Eq.~\eqref{eq:pseudounitarity}]~\cite{Globosits}. In the following, we show that there is no contradiction between these two observations. Importantly, when counting the number of pseudophotons, negative-frequency components have to be weighted with an additional minus sign via the matrix $V$ from Eq.~\eqref{eq:V_def}. For example, the number of pseudophotons of an arbitrary photon-flux normalized input light field reads $\expval*{V}{\instate}=\sum_n \sign(\omega_n)\abs*{\instate_n}^2$. Contrary, for the same state, the number of photons is given by $\braket*{\instate}=\sum_n \abs*{\instate_n}^2$. Distinguishing between these two quantities is essential for understanding what happens at EPs and at the CPA and lasing condition that are induced by time modulations.

To discuss this in detail, let us first consider eigenstates of $\SF$ in the broken symmetry regime $\ket*{\instate_\nupm}$ with eigenvalues $\abs*{\lambda_\nupm} \neq 1$. Here, the number of photons is not conserved but rather reduced (for~$\abs*{\lambda_\num}~<~1$) or enhanced (for~$\abs*{\lambda_\nup}~>~1$) during the scattering process as
\begin{equation}
\begin{split}
    \braket*{\outstate_\nupm} = \expval*{\SF^\dagger \SF}{\instate_\nupm} = \abs{\lambda_\nupm}^2 \braket*{\instate_\nupm} 
\end{split}
\end{equation}
The above reasoning also includes the extreme cases of perfect absorption $\abs*{\lambda_\num}=0$ and lasing $\abs*{\lambda_\nup}=\infty$, respectively. Interestingly, eigenstates in the broken regime are characterized by a vanishing number of pseudophotons: By considering the pseudounitarity condition [Eq.~\eqref{eq:pseudounitarity}] for states in the broken regime, we arrive at
\begin{equation}
\begin{split}
   \expval*{V}{\instate_\nupm} = \abs{\lambda_\nupm}^2 \expval*{V}{\instate_\nupm}
\end{split}
\end{equation}
As $\abs*{\lambda_\nupm}^2 \neq 1$, this entails $\expval*{V}{\instate_\nupm}=0$. Such a condition can only be achieved if the eigenstates $\ket*{\instate_\nupm}$ have an equal amount of photons associated with their negative- and positive-frequency channels. This result highlights the importance of the interplay between positive- and negative-frequency components for states in the broken regime. Crucially, however, the fact that the number of pseudophotons of the states $\ket*{\instate_\nupm}$ vanishes does not necessarily mean that the photon content also vanishes. Rather, such states represent light fields that can carry a finite amount of energy as $\braket*{\instate_\nupm}$ can be finite even though $\expval*{V}{\instate_\nupm}=0$.

On the contrary, there exist states that not only conserve the number of pseudophotons during scattering, but additionally they also conserve the number of photons. One example are eigenstates of $\SF$ in the unbroken symmetry regime as 
\begin{equation}
\begin{split}
    \braket*{\outstate_\nuz} = \abs{\lambda_\nuz}^2 \braket*{\instate_\nuz} = \braket*{\instate_\nuz} 
\end{split}
\end{equation}
In general, however, a superposition of states from the unbroken regime does not conserve the number of photons due to the non-orthogonality of the eigenstates of $\SF$. We provide further details on the conservation of pseudophotons, photons, and energy in Sec.~\ref{app:photonconserv} of the Supplementary Information.

Another example of states that conserve the number of photons during scattering arises if positive- and negative-frequency components of the light field do not mix during the scattering process (for example, for weak and slow modulations). In this case, the pseudounitarity condition of $\SF$ reduces t o a unitarity condition for the positive- and negative-frequency components individually. Then, eigenstates of the Floquet scattering matrix are orthogonal, and any superposition of such eigenstates conserves the number of photons during scattering. In such a case, no symmetry breaking and no EPs arise, preventing also the scenarios of CPA and lasing.
 
\subsection{Complex Scattering Objects}
\begin{figure*}
\includegraphics[width= .95\textwidth]{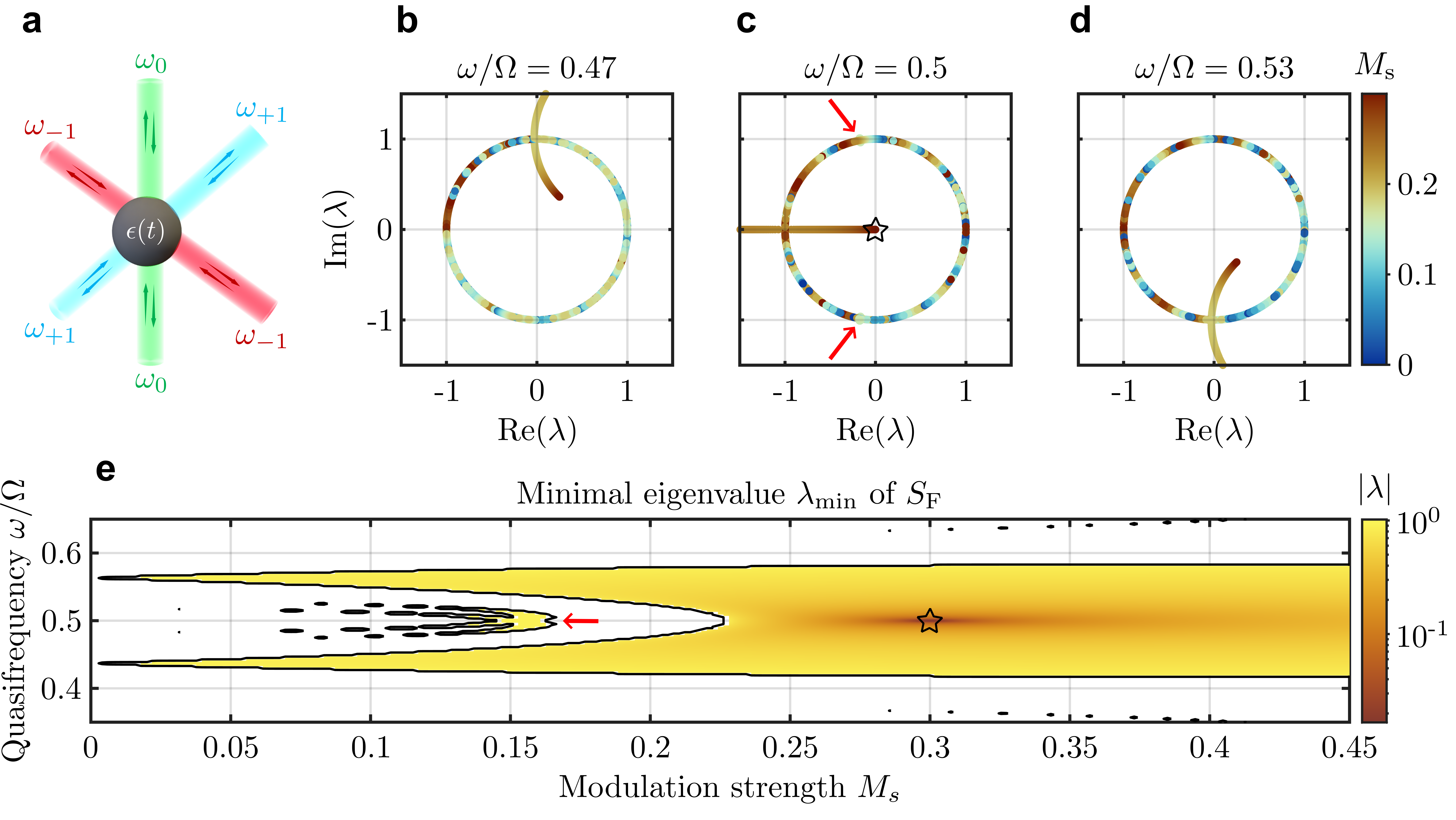}
\caption{Behavior of the eigenvalues $\lambda$ of the Floquet scattering matrix $\SF$ for a time-varying isolated sphere (see text for parameter values). \textbf{a},~We consider the properties of light scattering off a sphere with a time-periodic permittivity function. \textbf{b-d},~The complex eigenvalues of $\SF$ for varying modulation strengths $\ms$ and for different quasifrequencies $\omega/\Omega = 0.47, \, 0.5, \, 0.53$, respectively. For all three choices of the input quasifrequency, EPs form, and the system undergoes symmetry-breaking transitions. The red arrows indicate the appearance of additional EPs primarily associated with higher Floquet channels $n<-1$ and $n>0$. As the modulation strength increases, they enter the broken regime but quickly recombine on the unit circle again. Furthermore, only in \textbf{c}, where the light field fulfills the parametric resonance condition $\omega/\Omega=0.5$, some eigenvalues vanish. At this operational condition, the system can act as a CPA (star symbol). Since the eigenvalues come in inverse complex conjugate pairs, there simultaneously exist diverging eigenvalues corresponding to the system acting as a laser (not shown). \textbf{e},~The absolute value of the minimal eigenvalue of $\SF$ is shown here as a function of the quasifrequency $\omega$ and modulation strength $\ms$. The unbroken regime where all eigenvalues are unimodular (white) is separated from the broken regimes in which $\abs{\lambda_\mathrm{min}}<1$ (yellow) by exceptional lines (black lines). The star symbol indicates the CPA condition. The red arrow in panel~\textbf{e} indicates the broken regime associated with the additional EPs visible in panel~\textbf{c}.}
\label{fig:sphere}
\end{figure*}

\begin{figure}[h]
\includegraphics[width= 0.99\columnwidth]{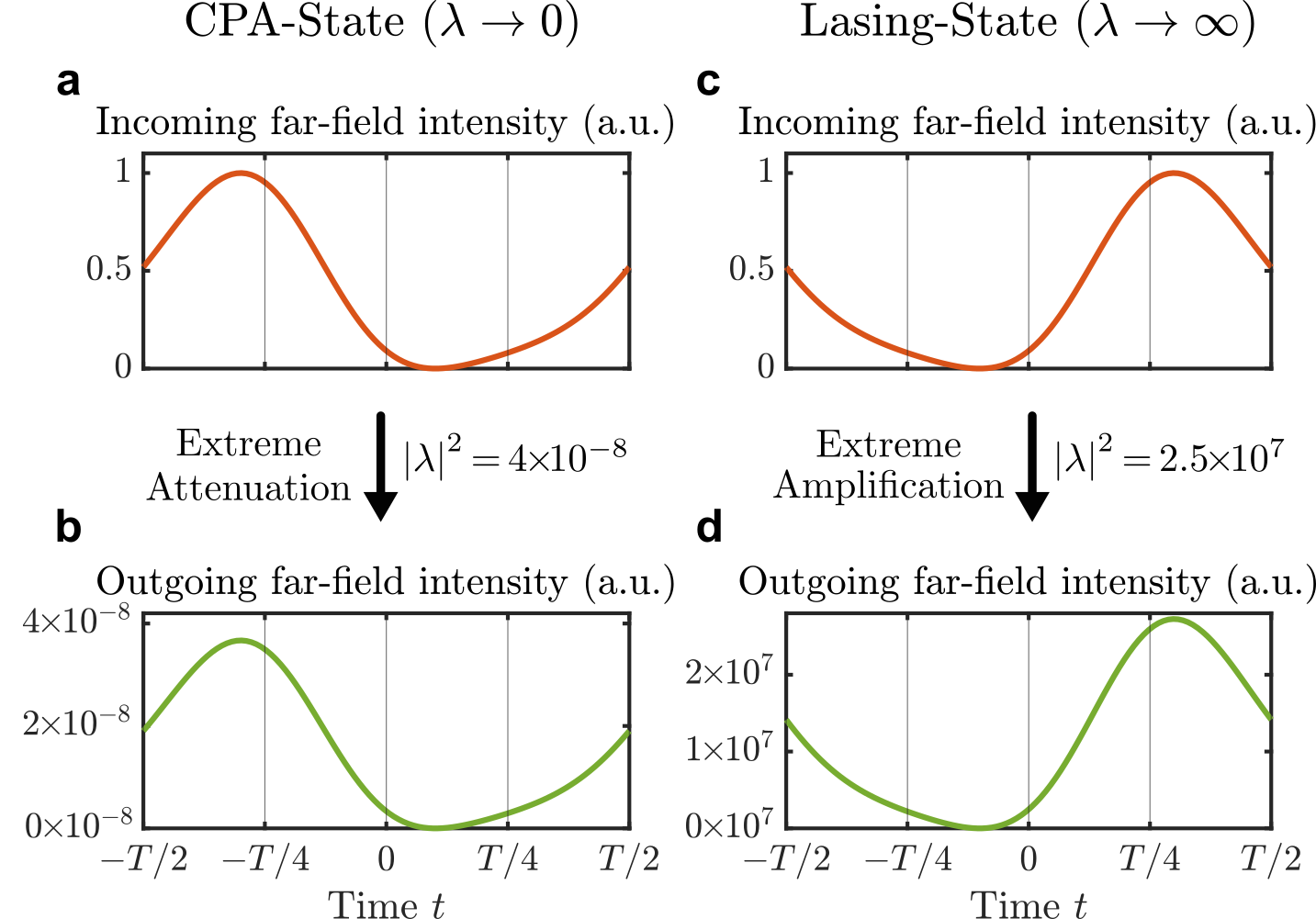}
\caption{Space-integrated incoming and outgoing intensities for the CPA and lasing states for the time-varying sphere (see text for parameter values) as a function of time $t$. We evaluate the electric fields in the far-field at $r=1000 \times Tc$ and spatially integrate them on an imaginary sphere that encloses the time-varying sphere. \textbf{a},~Incoming and \textbf{b},~outgoing intensity corresponding to the CPA state with an eigenvalue $\lambda=2\times 10^{-4}$. An extreme attenuation of the intensity can be observed, and nearly no outgoing field is produced. \textbf{c},~Incoming and \textbf{d},~outgoing intensity of the lasing state corresponding to an eigenvalue $\lambda=5\times 10^{3}$. Here, we observe an extreme amplification of the incoming light field. Furthermore, we see the time-reversal symmetry of the CPA and lasing states: The incoming CPA state is the time-reversed of the outgoing lasing state.}
\label{fig:intensity_sph}
\end{figure}

In this section, we demonstrate that our formalism is not only applicable to the basic one-dimensional slab system but also to much more complex scattering objects. Specifically, we study an isolated time-varying sphere and a metasurface made from these spheres.

First, we consider the scattering properties of an isolated time-varying sphere surrounded by free space, thus constituting a localized scatterer in three-dimensional space [see Fig.~\ref{fig:sphere}a]. The permittivity $\epsilon(\mathbf{r},t)$ of the system is given by 
\begin{equation}
    \epsilon(\mathbf{r},t)=1+\chiz [1 + \ms \cos(\Omega t)]\Theta(R-{r})\label{eq:epsilon_sph}
\end{equation}
We set $\chiz=11.67$, which corresponds to the susceptibility of silicon at near-infrared frequencies. Furthermore, we choose the radius of the sphere to be $R=568$~nm and set the modulation frequency to $\Omega=2\pi\times151$~THz. We compute the $\SF$ matrix of the time-varying sphere following the Mie theory-based T-matrix method introduced in Ref.~\onlinecite{ptitcyn2023floquet} using $n \in [-8,7]$ Floquet channels and multipoles up to the octupole order (for details see Sec.~\ref{app:sphere} of the Supplementary Information). 

In Fig.~\ref{fig:sphere}b-d, we plot all complex eigenvalues of $\SF$ for this system as a function of the modulation strength $\ms$ and for three choices of the input quasifrequency: $\omega/\Omega=0.47,\,0.5,\,0.53$. Similar to the slab system, we observe in all three cases that for a weak modulation, i.e., $\ms \ll 0.1$, all eigenvalues reside on the unit circle. As $\ms$ increases, EPs form, signaling the onset of the broken regime. In Fig.~\ref{fig:sphere}b and d, we find that if the wave field is detuned from the parametric resonance condition, the eigenvalues that enter the broken symmetry regime do not approach $0$ or $\infty$. Only for $\omega/\Omega=0.5$ we see that there exist eigenvalues for which $\abs{\lambda} \rightarrow 0$ [see Fig.~\ref{fig:sphere}c] and, therefore, also eigenvalues $1/\abs{\lambda^*} \rightarrow \infty$ for the same $\ms$ (not shown). This confirms that, like in the slab setup, a sphere made from a material with a periodically time-varying permittivity can also simultaneously act as a CPA and as a laser. We find that the eigenvalues corresponding to the CPA and lasing points arise primarily due to the magnetic dipolar part of the $\SF$ matrix of the sphere. Furthermore, due to the symmetry of the sphere along all three spatial dimensions, these eigenvalues have a three-fold degeneracy. Note that we observe additional EPs [indicated by red arrows in Fig.~\ref{fig:sphere}c and e], which only appear due to the higher-order Floquet channels ($n<-1$ and $n>0$). However, for the parameter values we consider here, these eigenvalues do not reach the CPA and lasing condition, but rather recombine at the unit circle again for increasing modulation strength.

As for the slab, we investigate the influence of the quasifrequency $\omega$ and of the driving strength $\ms$ on the eigenvalues of $\SF$ also in the present case. The corresponding plot of the minimal eigenvalue $\lambda_\mathrm{min}$ of $\SF$ as a function of $\omega$ and $\ms$ is shown in Fig.~\ref{fig:sphere}e. If the incident light field is spectrally far detuned from the parametric resonance condition $\omega/\Omega=0.5$, no exceptional points are formed (white region). However, near $\omega/\Omega = 0.5$, the system can undergo a symmetry-breaking transition and EPs appear (black curve). By choosing $\omega/\Omega = 0.5$ and $\ms= 0.3$, we observe the CPA-lasing point indicated by a nearly vanishing eigenvalue. Specifically, we numerically find that for this set of parameters, $\abs{\lambda_\mathrm{min}}=2 \times 10^{-4}$. To prove that in the vicinity of this small eigenvalue, there truly exists a vanishing eigenvalue, we again calculate the winding number over a loop enclosing this CPA point. We find $\wind(\SF) \approx 6-3.5\times10^{-4}$, as expected due to the three-fold degeneracy of each eigenvalue.
Next, we investigate the spatiotemporal behavior of the associated CPA and lasing states. In Fig.~\ref{fig:intensity_sph}, we plot the space-integrated incoming and outgoing far-field intensity corresponding to the eigenstates of the minimal (CPA) and maximal (lasing) eigenvalues. For numerical convenience, the intensity is space-integrated on the surface of an imaginary sphere with radius $r=1000 \times Tc$ ($T=2\pi/\Omega$ and $c$ is the speed of light in vacuum), such that $r \gg R$. For the CPA state, we observe an extreme reduction of the outgoing wave intensity compared to the input [see Fig.~\ref{fig:intensity_sph}a-b]. Conversely, we see that if we choose the lasing state as the input, the outgoing intensity is greatly enhanced, as expected [see Fig.~\ref{fig:intensity_sph}c-d]. Furthermore, we can observe the time-reversal symmetry of both states according to Eq.~\eqref{eq: time-reversal}: the input CPA field is the time-reversed of the output lasing field.

\begin{figure*}[htbp]
\includegraphics{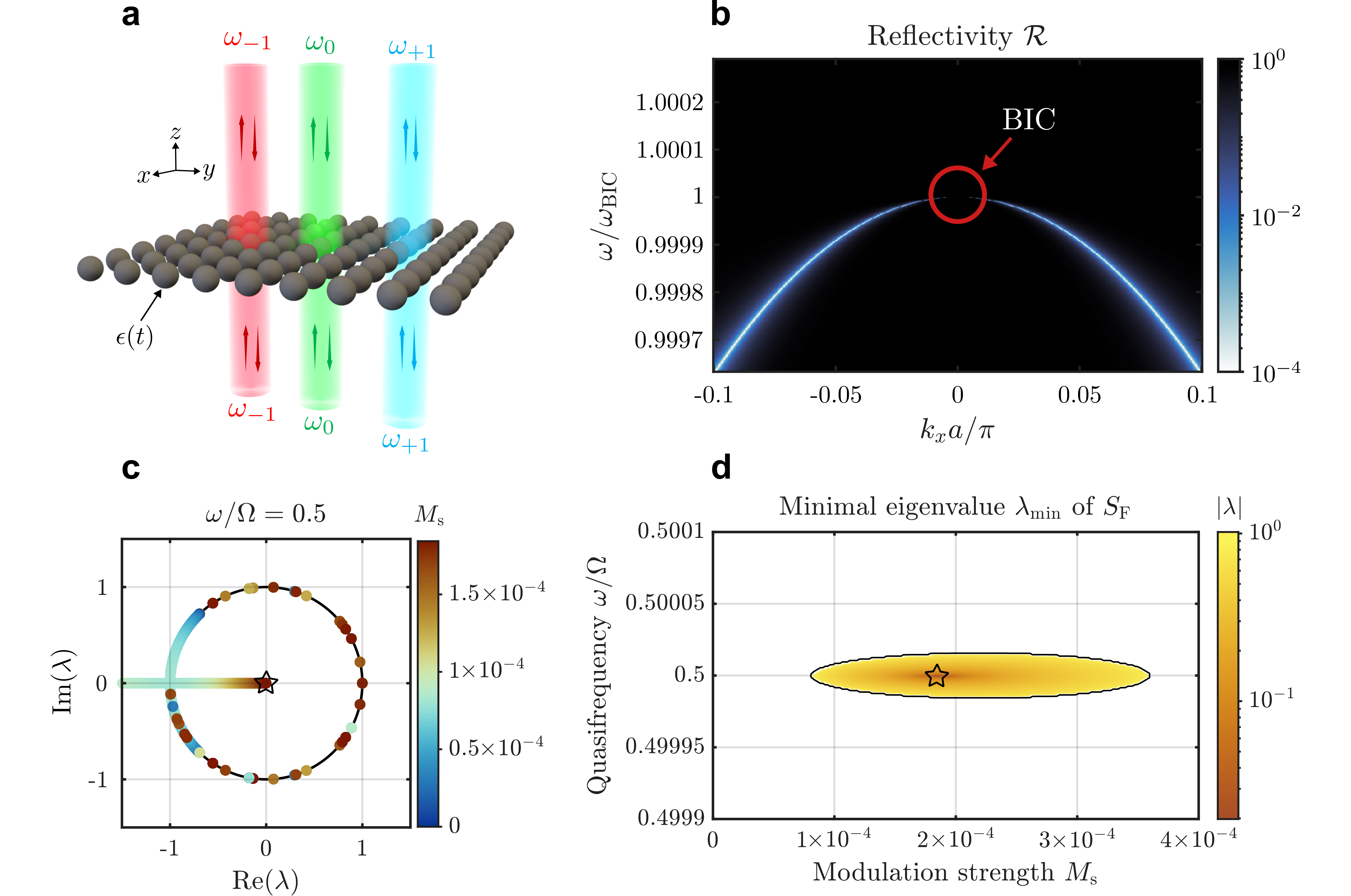}
\caption{Behavior of the eigenvalues $\lambda$ of the Floquet scattering matrix $\SF$ for a time-varying metasurface. \textbf{a},~We consider the properties of light scattering off a metasurface that consists of a periodic arrangement (on a square lattice) of spheres made from a time-varying medium. The time-varying medium is characterized by the permittivity $\epsilon(\mathbf{r},t)=1+\chiz [1 + \ms \cos(\Omega t)]f(\mathbf{r})$. Here, $f(\mathbf{r})=1$ for $\mathbf{r}$ at the spatial domains occupied by the spheres and $0$ otherwise (see text for the other parameter values). \textbf{b}, The reflectivity $\mathcal{R}$ of the metasurface shown in \textbf{a} as a function of $\omega$ and the $x$-component of the Bloch wavevector $k_x$ under static conditions (i.e., $\ms=0$) and $k_y=0$. We note that at $k_x=0$, there exists a BIC that has a symmetry compatible with TE-polarized plane waves. \textbf{c},~The complex eigenvalues of $\SF$ for varying modulation strengths $\ms$ and for the quasifrequency $\omega/\Omega = 0.5$. Here, we observe that at a certain $\ms$, the light field fulfills the parametric resonance condition, leading to vanishing eigenvalues (indicated by the star symbol), signifying CPA and lasing. \textbf{d},~The absolute value of the minimal eigenvalue of $\SF$ as a function of the quasifrequency $\omega$ and modulation strength $\ms$. The unbroken regime where all eigenvalues are unimodular (white) is separated from the broken regime in which $\abs{\lambda_\mathrm{min}}<1$ (yellow) by an exceptional line (black). The star symbol indicates the CPA condition. To prove that $\abs{\lambda_\mathrm{min}}$ vanishes entirely at the CPA point, we calculated the winding number over a loop enclosing the CPA point. We find that the winding number is $\wind(\SF) \approx 2-10^{-4}$, as expected, due to the TE-polarized qBIC mode of the metasurface.}
\label{fig:metasurface}
\end{figure*}

We already saw that by exploiting static resonances we can reduce the $\ms$ needed to form EPs (see Subsec.~\ref{app:tbm_EP}). Now, using high-quality factor (high-Q) resonances, we show that we can also reduce the modulation strength $\ms$ required for CPA and lasing points~\cite{garg2025photonic}. In particular, we use the quasi-bound states in the continuum (qBICs) of metasurfaces, which are not present in the slab (and homogeneous sphere) system~\cite{garg2025photonic,hsu2016bound}. The considered metasurfaces are made from a periodic arrangement on a square lattice of spheres made from a time-varying medium [see Fig.~\ref{fig:metasurface}a]. 

First, we consider the case of a static metasurface without any time-modulation, i.e., $\ms=0$. We optimize the geometry parameters of the metasurface such that it supports a bound state in the continuum (BIC) at normal incidence, i.e., for the Bloch wavevector $\mathbf{k}_\parallel=(k_x,k_y)^\mathrm{T}=0$~\cite{sadrieva2019multipolar,Ustimenko2024resonances}. The radius of the spheres in the metasurface is $R=225$~nm, and the lattice constant is $a=3R$. The reflectivity $\mathcal{R}$ of the metasurface is shown in Fig.~\ref{fig:metasurface}b as a function of $\omega$ and the $x$-component of the Bloch wavevector $k_x$. For simplicity, we assume $k_y=0$. Here, we observe a sharp resonant behavior of $\mathcal{R}$ in the $\omega-k_x$ space. As $k_x\rightarrow0$, the linewidth of the resonance becomes vanishingly small, and the resonance eventually disappears at $k_x=0$ and $\omega=\omega_\mathrm{BIC}$, indicating a BIC there, which does not couple to external radiation. In the following, we use a qBIC of the metasurface, which is formed by slightly detuning the parameters of the system from the perfect BIC condition, such that the resulting resonance has a high but finite Q-factor.

For the simulations of the time-varying metasurface, we choose $k_x=0.05\pi/a$  and correspondingly $\omega=\omega_\mathrm{qBIC}=0.99991\omega_\mathrm{BIC}$. Note that the numerical value of $\omega_\mathrm{qBIC}$ has to be chosen precisely due to the extremely narrow linewidth of the qBIC [see Fig.~\ref{fig:metasurface}b]. Furthermore, we use $\Omega=2\omega_\mathrm{qBIC}$, which ensures that upon time-modulation, the modes of the static metasurface at $\omega_0=\omega_\mathrm{qBIC}$ and $\omega_{-1}=-\omega_\mathrm{qBIC}$ are resonantly coupled to each other at $\Omega=2\pi\times387$~THz, such that the resulting hybrid modes have minimal radiation losses. We numerically compute the corresponding Floquet scattering matrix of the time-varying metasurface following the Mie theory-based T-matrix method introduced in Refs.~\onlinecite{garg2022modeling,garg2025photonic} using $n \in [-2,1]$ Floquet channels and multipoles up to the octupolar order (see Sec.~\ref{app:metasurface} in the Supplementary Information for details). The metasurface is subwavelength at the frequencies $\omega_0$ and $\omega_{-1}$, and the first diffraction order is propagating for the frequencies $\omega_{-2}$ and $\omega_{1}$. We verified that this choice of parameters is sufficient to ensure numerical convergence.

To show the existence of EPs, CPA, and lasing points, we plot all the complex eigenvalues $\lambda$ of the corresponding $\SF$ as a function of $\ms$ for the quasifrequency $\omega/\Omega =0.5$ in Fig.~\ref{fig:metasurface}c. As predicted by our formalism, we observe that as $\ms$ is increased, the eigenvalues coalesce on the unit circle forming an EP. On a further increase of $\ms$, these eigenvalues leave the unit circle in a pairwise manner. We also find that at a certain $\ms$, there exists an eigenvalue whose absolute value approaches $0$, marking a CPA point. Conversely, for its inverse conjugate pair $1/\lambda^*$, the absolute value diverges for the same driving strength $\ms$ (not shown). However, note that contrary to the previous scattering setups, the modulation strengths necessary to form EPs, and to make the system operate as a CPA or laser, is drastically reduced. In particular, to form an EP, we require $\ms=8.16\times10^{-5}$, and for the CPA/lasing point, we require $\ms=1.84\times10^{-4}$. In theory, by moving the operation point closer to the BIC, i.e., $\omega\rightarrow\omega_\mathrm{BIC}$ and $k_x\rightarrow0$, the modulation strength required for EPs, CPA, and lasing points can be made arbitrarily small, i.e., $\ms\rightarrow0$. However, in practice, unavoidable fabrication imperfections and material losses of realistic structures lower the Q-factor of the qBICs, leading to a lower bound on threshold $M_\mathrm{s}$ to reach EPs, CPA, and lasing points. Nevertheless, structures supporting high-Q qBICs serve as a promising platform to achieve low-threshold lasing~\cite{Khajavikhan}. To highlight the CPA and lasing conditions more clearly, we plot the minimal eigenvalue $\lambda_\mathrm{min}$ of $\SF$ of the metasurface as a function of $\omega$ and $\ms$ in Fig.~\ref{fig:metasurface}d. Here, we observe that at $\omega / \Omega=0.5$ and $\ms=1.84\times10^{-4}$, the system behaves as a CPA indicated by $\abs{\lambda_\mathrm{min}} \rightarrow 0$. Furthermore, as predicted by our theory, for the same values of $\omega$ and $\ms$, $\abs{\lambda_\mathrm{max}} \rightarrow \infty$, marking that the system acts as a laser there. We refer the reader to Sec.~\ref{app:MS_int} of the Supplementary Information for details on the symmetry of the far-field intensities assuming the CPA and lasing eigenstates as input fields.

\section{Discussion}
We have shown that in periodically time-varying systems of finite size, symmetry-breaking transitions occur at exceptional points. Specifically, we revealed that the eigenvalues of the Floquet scattering matrix are either unimodular, corresponding to the unbroken symmetry regime, or come in inverse complex-conjugate pairs, corresponding to the broken symmetry regime. This behavior of the eigenvalues is explained through the pseudounitarity of the Floquet scattering matrix. We provided an expression for the associated symmetry operator for arbitrary temporal modulations and showed that for a time-symmetric drive, this operator is the time-reversal operator. Furthermore, we demonstrated that if the incident light field has a quasifrequency of half the modulation frequency and thus fulfills the parametric resonance condition, a special situation can occur in which one eigenvalue vanishes while another one diverges. At this point, the system simultaneously acts as a CPA and as a laser. We demonstrated the working principle of our formalism using the example of a time-varying slab, a time-varying isolated sphere, and a time-varying metasurface. Leveraging the qBICs of metasurfaces, the modulation strength needed to achieve CPA and lasing points can be made arbitrarily small.

Our framework applies to objects of arbitrary shape. We expect it to be relevant across different experimental platforms, where the characteristics of individual systems can be appropriately taken into account. This includes, but is not limited to, PTCs in the optical regime based on epsilon-near-zero materials~\cite{Tirole,Lustig23}, in the microwave regime based on an LC resonator setup~\cite{Hooper,tuxbury2025synthetic}, or using water waves~\cite{Bacot,Apffel}. In particular, in the optical regime, CPA and lasing behavior has been observed experimentally for a modulation frequency $\Omega\approx2\pi\times 512$~THz and strength $M_\mathrm{s}\approx0.016$ of an indium tin oxide (ITO) slab \cite{Galiffi24}. Our work also provides a bridge to the quantum optics domain~\cite{Sustaeta25}, where it will be of interest to study the implications of our scattering theory for the dynamical Casimir effect~\cite{Maghrebi,Dodonov} or for the properties of the proposed Floquet laser (photon statistics, coherence properties etc.)~\cite{Haken}.

\section{Materials and methods}
\subsection{Reciprocity} \label{sec:reciprocity}
Here, we provide further insights into the connection between time-reversal symmetry and the Floquet scattering matrix. Specifically, we show that for time-reversal symmetric Floquet scattering systems described by $\epsilon(t)=\epsilon(-t)$, the Floquet scattering matrix can be chosen to satisfy $V\SF^\mathrm{T}V=\SF$. To achieve this, we first prove that $\SF^* = \SF^{-1}$ holds in the considered case. Then, using the pseudounitarity condition Eq.~\eqref{eq:pseudounitarity}, we can immediately conclude that $V\SF^\mathrm{T}V=\SF$. To keep the derivation general, we expand the far-field electric field as $\vb{E}(\vb{r},t) = \sum_n \vb{E}_n(\vb{r}) e^{- i\omega_n t}$, where $ \vb{E}_n(\vb{r}) = \sum_{\alpha} \cin_{n,\alpha} \vb{Z}_{n,\alpha}(\vb{r}) + (\SF \cinv)_{n,\alpha} \vb{Z}_{n,\alpha}^*(\vb{r})$. Here, $\vb{Z}_{n,\alpha}(\vb{r})$ are the incoming far-field spatial mode functions labeled by $\alpha$ and $\vb{Z}^*_{n,\alpha}(\vb{r})$ are the corresponding outgoing far-field modes. Furthermore, we used that $\coutv = \SF \cinv$. Next, we complex conjugate the outgoing field coefficients $\SF \cinv$ and reinsert them into the system such that the corresponding far field $\vb{\tilde{E}}(r,t)=\sum_n \vb{\tilde{E}}_n(\vb{r}) e^{- i\omega_n t}$ is given by $\tilde{\vb{E}}_n(\vb{r}) = \sum_{\alpha} (\SF\cinv)_{n,\alpha}^* \vb{Z}_{n,\alpha}(\vb{r}) + [\SF(\SF\cinv)^*]_{n,\alpha} \vb{Z}_{n,\alpha}^*(\vb{r})$. The crucial step now is to show that $\vb{E}(\vb{r},t)$ and $\vb{\tilde{E}}(\vb{r},t)$ represent time-reversed fields of each other. This can be seen by noting that $\vb{E}_n^*(\vb{r})$ and $\vb{\tilde{E}}_n(\vb{r})$ consist of the same incident field coefficients $(\SF\cinv)^*_n$ and are solutions to the same wave equation
\begin{eqnarray}
    \nabla \times (\nabla \times \vb{E}_n^*) &= k_n^2 \sum_m \epsilon_{n-m}^* k_m \vb{E}_m^* \\
    \nabla \times (\nabla \times \vb{\tilde{E}}_n) &= k_n^2 \sum_m \epsilon_{n-m} k_m \vb{\tilde{E}}_m 
\end{eqnarray}
since for time-symmetric modulations we have $\epsilon_{n-m}^* = \epsilon_{n-m}$. This implies that $\SF(\SF\cinv)^* = (\cinv)^*$, proving that $\SF^* = \SF^{-1}$.

\subsection{Symmetry Operators \label{app:symmetry_operators}}
In this subsection, we provide details on how the symmetry operator $X$ introduced in Eq.~\eqref{eq:X_definition} can be constructed for diagonalizable pseudounitary Floquet scattering matrices. The following discussion is general in the sense that we do not assume any spatial or temporal symmetries. We start by noting that, by definition, the right eigenvectors $\ket*{\evecR_{\nu}}$ and the left eigenvectors  $\ket*{\evecL_{\nu}}$ of the Floquet scattering matrix $\SF$ satisfy
\begin{eqnarray}
    \SF \ket*{\evecR_{\nu}} &= \lambda_{\nu} \ket*{\evecR_{\nu}}  \\
    \SF^\dagger \ket*{\evecL_{\nu}} &= \lambda_{\nu}^* \ket*{\evecL_{\nu}} 
\end{eqnarray}
Furthermore, we observe that the matrix $V$ transforms a right eigenvector into a left eigenvector, which is a direct consequence of the pseudounitary relation Eq.~\eqref{eq:pseudounitarity}, as
\begin{equation}
    \SF^\dagger V \ket*{\evecR_{\nu}} = V \SF^{-1} \ket*{\evecR_{\nu}} = \frac{1}{\lambda_{\nu}} V \ket*{\evecR_{\nu}}  \label{eq:Vpsi}
\end{equation}
such that we can read off $\ket*{\evecL_{\mu}}=V \ket*{\evecR_{\nu}}$. Furthermore, left and right eigenvectors can be chosen to be biorthonormal
\begin{equation}
    \braket*{\evecL_{\mu}}{\evecR_{\nu}} = \delta_{\mu,\nu} 
\end{equation}

Here, we closely follow~\cite{Mostafazadeh03}. Using left and right eigenvectors, we can now construct an anti-linear symmetry operator as
\begin{equation}
    X = \sum_\nuz \ket*{\evecR_\nuz} \star \bra*{\evecL_\nuz} + \sum_\nupm \ket*{\evecR_\nup}\star \bra*{\evecL_\num} + \ket*{\evecR_\num}\star \bra*{\evecL_\nup} 
\end{equation}
Here, the symbol $\star$ represents complex conjugation in the following way
\begin{equation}
    \star \braket*{a}{b} = \braket*{a}{b}^* = \braket*{b}{a} 
\end{equation}
By definition, the operator $X$ satisfies
\begin{equation}
    X \ket*{\evecR_{\nu}} =
    \begin{cases}
        \ket*{\evecR_{\nuz}}, \quad & \nu=\nuz \\
        \ket*{\evecR_{\nupm}}, \quad & \nu=\nump
    \end{cases}
\end{equation}
and thus constitutes a symmetry operator that maps states from the symmetry-unbroken regime onto themselves and states in the symmetry-broken regime onto their partner state.

\subsubsection{Degeneracies}
Due to the underlying spatial symmetries of the scattering system, the associated Floquet scattering matrix $\SF$ may have degenerate eigenvalues. Suppose $\SF$ has a $J$-fold degenerate eigenvalue $\lambda$ to the corresponding normalized eigenvectors $\ket*{\evecR_{j}}$ for $j\in[1,J]$. This degeneracy implies that any linear combination of the eigenvectors $\ket*{\evecR_{j}}$ is also an eigenvector of $\SF$ with the same eigenvalue $\lambda$. Therefore, there are many sets of normalized eigenvectors of $\SF$, and Eq.~\eqref{eq:X_definition} cannot be applied to every such set. However, as verified numerically, there always exists at least one set of eigenvectors of $\SF$ to which Eq.~\eqref{eq:X_definition} is applicable.

\subsection{Two-Band Model \label{app:tbm}}
Here, we provide additional details on the formation of EPs and the operational conditions for CPA and lasing for the time-varying slab. We employ the two-band model (TBM) approximation to arrive at an analytic result for the minimal modulation strength required for the formation of an EP. This enables us to investigate the impact of the normalized thickness of the slab on the formation of EPs. We analytically demonstrate that for specific slab thicknesses, even a very weak (infinitesimal) driving strength is sufficient to observe an EP. Furthermore, using a resonant state expansion (RSE), we provide an approximate analytical expression for the minimal modulation strength required for the system to act as a CPA or as a laser, solely based on the scattering parameters of the corresponding static system.

\subsubsection{Minimal $\ms$ for an EP \label{app:tbm_EP}}
 \begin{figure}
    \includegraphics[width=\columnwidth]{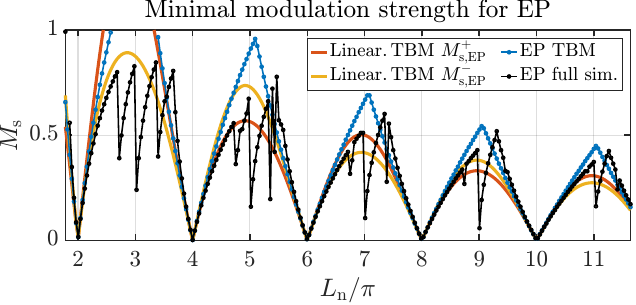}
    \caption{\label{fig:slab_EP_modulation} Minimal modulation strength $\depsEPmin$ necessary to observe an EP for the time-varying slab as a function of its normalized thickness $\Ln$ for $\omega/\Omega=0.5$. The red and yellow lines represent the results derived from the linearized TBM $\depsEP^{+}$ (red) and $\depsEP^{-}$ (orange). Using the TBM without an additional linearization, we find that EPs form at the same modulation strength (blue line). The black line corresponds to data from a full simulation including multiple frequency components $n \in [-8,7]$. In all cases, we see that for $\Ln=2m\pi$ with $m\in\mathbb{N}$ only a minimal modulation is needed for an EP to form, as analytically predicted by the linearized TBM.}
\end{figure}

In this subsection, we want to gain further insights into the formation of exceptional points and derive an expression for the minimal driving strength necessary to observe an EP [see~Eq.~\eqref{eq:deps_EP}]. For this, we fix the quasifrequency to $\omega/\Omega=1/2$ since we expect EPs to appear at lower modulation strengths for wave fields obeying the parametric resonance condition. Furthermore, we assume that only two modes ($n=-1,0$) are necessary to describe the wave field inside the modulated slab (TBM approximation) ~\cite{Asadchy}. The Maxwell equations inside the slab reduce in this case to
\begin{equation}
    \frac{\Omega^2}{4}\mqty( 1+\chiz & \chiz \ms / 2 \\
           \chiz \ms /2  & 1+\chiz) \mqty(e_{-1}\\e_{0}) = q^2 c^2 \mqty(e_{-1}\\e_{0}) 
\end{equation}
Solving the above eigenvalue problem results in wavevectors
$q_{p}=\frac{\Omega}{2c}\sqrt{1+\chiz \pm \chiz\ms/2}$ and eigenvectors $\vb{e}_p=(\pm 1, 1)^\mathrm{T}$ with band index $p=\{1,2\}$. We thus have an analytical (approximate) solution for the wave field within the PTC-slab. We utilize this solution to establish the corresponding Floquet scattering matrix. To arrive at an analytic model, we linearize the above result and retain only terms up to linear order in $\ms$. The resulting $4 \times 4$ Floquet scattering matrix takes the symmetric form
\begin{equation}
    \SFt=\mqty(\rFt & \tFt \\ \tFt & \rFt) 
\end{equation}
where expressions with a tilde stand for results derived using the approximations described above. We find the eigenvalues $\tlambda$ by making use of
\begin{equation}
\begin{split}
    0&=\det(\SFt-\tlambda)=\det \mqty(\rFt-\tlambda & \tFt \\ \tFt & \rFt-\tlambda) \\
    &=\det(\rFt+\tFt-\tlambda)\det(\rFt-\tFt-\tlambda) 
\end{split}
\end{equation}
Such a finding implies that each of the $2 \times 2$ matrices $\rFt\pm\tFt$ contributes a pair of eigenvalues, which we label as $\tlambda^\pm_{1,2}$. We, therefore, expect two EPs: one originating from the coalescence of $\tlambda^+_1$ and $\tlambda^+_2$; and one from the coalescence of $\tlambda^-_1$ and $\tlambda^-_2$. We analytically find that the exceptional points form at the critical driving strengths
\begin{equation}
    \depsEP^{\pm}=\frac{4(1+\chiz)\sqrt{1-\cos(\Ln)}}{\chiz \sqrt{1+\Ln^2/2-\cos(\Ln) \pm 2\Ln\sin(\Ln/2)}} \label{eq:deps_EP_full}
\end{equation}
with the normalized thickness $\Ln = L\Omega\sqrt{1+\chiz}/c$. Equation~\eqref{eq:deps_EP_full} is the complete version of Eq.~\eqref{eq:deps_EP}.

In Fig.~\ref{fig:slab_EP_modulation}, we now compare $\depsEP^{\pm}$ (red and yellow lines) to numerical data to investigate the validity of Eq.~\eqref{eq:deps_EP_full}, where we made use of several approximations. For comparison, we calculate the minimal modulation strength necessary to observe an EP again, considering only the two modes $n=-1,0$, but without linearizing with respect to the modulation strength (blue line). We find that, contrary to the prediction of the linearized TBM, both EPs form at the same modulation strength, which is why we only plot a single blue line. Furthermore, we present simulated data involving several frequency harmonics $n \in [-8,7]$ (black line). In all cases, we observe that, indeed, in the vicinity of $\Ln=2 m\pi$ with $m\in \mathbb{N}$ only a very weak modulation is necessary to realize an EP as predicted by Eq.~\eqref{eq:deps_EP_full}, respectively. One feature, which we only observe using the full simulation, is the occurrence of several dips. They primarily appear when the TBM predicts large modulation strengths for the formation of an EP [around $\Ln=(2m+1)\pi$]. These dips correspond to EPs, corresponding to a symmetry breaking of modes that are associated with higher Floquet channels $n>0$ and $n<-1$. They are thus only visible in a description that goes beyond the TBM using more than two channels.

\subsubsection{Minimal $\ms$ for CPA and Lasing \label{app:tbm_CPA}}
 \begin{figure}
    \includegraphics[width=0.99\columnwidth]{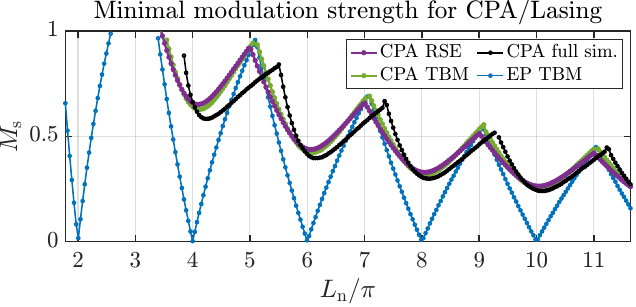}
    \caption{\label{fig:slab_CPA_modulation} Minimal modulation strength $\msCPAmin$ necessary such that the time-varying slab operates as a CPA or laser as a function of the normalized thickness $\Ln$ for $\omega/\Omega=0.5$. The purple line shows the result from the resonant state expansion (RSE) Eq.~\eqref{eq:msCPAmin}. The green line represents the results derived from an analysis of the eigenvalues of $\SF$ using the TBM approximation, showing strong agreement with the RSE approach, as expected. For reference, we also plot $\depsEPmin$ using the TBM approximation. We perform a full-scale simulation using $n \in [-8,7]$ Floquet channels (black line).}
\end{figure}

In the following, we derive an expression for the minimal driving strength necessary to observe CPA and lasing from the time-varying slab. Here, we make use of the Floquet resonant state theory developed in Ref.~\onlinecite{Valero25}. Corresponding to the TBM, we only take into account two static resonant states: First, we have $\rs{\alpha}$ with complex eigenfrequency $\rsomc{\alpha}=\rsom{\alpha}-i \rsgamma{\alpha}$, where $\alpha=0,1,\dots$ labels the resonant states. Secondly, we use the $m=-1$ replica of the negative twin with eigenfrequency $\Omega-\rsomc{\alpha}^*$. Here, $\rsom{\alpha} \in \mathbb{R}$ is the resonance frequency and $\rsgamma{\alpha} \in \mathbb{R}$ is the decay rate of the corresponding resonant state of the static system. Explicitly, for a slab of thickness $L$ and time-invariant permittivity $\epsilon=1+\chiz$ we have $\rsom{\alpha} = \frac{\alpha\pi c}{\sqrt{\epsilon}L}$ and $ \rsgamma{\alpha} = \frac{c}{\sqrt{\epsilon}L} \ln(\abs{1+\sqrt{\epsilon}}/\abs{1-\sqrt{\epsilon}})$~\cite{Lalanne,Weiss}. The resonant states inside the slab $\abs{x} \leq L/2$ read
\begin{equation}
        \rs{\alpha}(x) = 
    \frac{1}{\sqrt{\epsilon \varepsilon_0 L}} \begin{cases}
        \sin(\sqrt{\epsilon} \rsomc{\alpha} x / c)\,, \quad \text{if $\alpha$ is odd} \\
        \cos(\sqrt{\epsilon} \rsomc{\alpha} x / c)\,, \quad \text{if $\alpha$ is even}
    \end{cases}
\end{equation}
We introduce the detuning $\Delta_{\alpha}=\abs{\rsom{\alpha}-\Omega/2}$ and define the intensity of the resonant state inside the slab as $I_\alpha~=~\int_{-L/2}^{L/2} \dd{x} \abs{\rs{\alpha}(x)}^2$ and furthermore $g_{\alpha}~=~\chiz \ms I_\alpha/2$. In the following, we always choose that resonant state, i.e., that specific value of $\alpha$, corresponding to the resonant state that has minimal detuning $\Delta_{\alpha}$. Performing a similar analysis as in Ref.~\onlinecite{Valero25}, we find the eigenfrequencies of the Floquet resonant states
\begin{equation}
    \tilde{\omega}_\pm \approx \frac{\Omega}{2} - i\rsgamma{\alpha} \pm i g_{\alpha} \sqrt{\frac{\Omega^2}{4}-\frac{\Delta_{\alpha}^2}{g_{\alpha}^2}+\rsgamma{\alpha}^2} \label{eq:RS_disp}
\end{equation}
Crucially, a scattering system acts as a laser (and thus, in our case, simultaneously as a CPA) when the eigenfrequency of a Floquet resonant state has a vanishing imaginary part. In our case, the eigenfrequency $\tilde{\omega}_+$ becomes real when two conditions are fulfilled: (i) the discriminant in the square root of Eq.~\eqref{eq:RS_disp} is positive, and (ii) when the square root compensates for the decay rate of the original mode, such that
\begin{equation} \label{eq:comp}
    \rsgamma{\alpha} = g_{\alpha}\sqrt{\frac{\Omega^2}{4}-\frac{\Delta_\alpha^2}{g_{\alpha}^2}+\rsgamma{\alpha}^2}
\end{equation}
For the modes of interest, it typically holds that the decay rate is much smaller than the resonance frequencies. Then, Eq.~\eqref{eq:comp} becomes
\begin{equation} \label{eq:comp2}
    \begin{split}
        \rsgamma{\alpha} &= g_{\alpha}\sqrt{\frac{\Omega^2}{4}-\frac{\Delta_{\alpha}^2}{g_{\alpha}^2}} \\
        &= \frac{\chiz \ms I_{\alpha}\Omega}{4}\sqrt{1-\left(\frac{4\Delta_{\alpha}}{\chiz \ms I_{\alpha}\Omega}\right)^2}
    \end{split}
\end{equation}
Condition (i) tells us that $\frac{4\Delta_{\alpha}}{\chiz \ms I_{\alpha}\Omega}<1$. We therefore expand Eq.~\eqref{eq:comp2} according to $\sqrt{1-x^2} = 1-x^2/2+\mathcal{O}(x^4)$ valid for small $x$ and arrive at
\begin{equation} \label{eq:thr}
\rsgamma{\alpha} \approx \frac{\chiz \ms I_{\alpha}\Omega}{4} - \frac{2\Delta_{\alpha}^2}{\chiz \ms I_{\alpha}\Omega} 
\end{equation}
Solving Eq.~\eqref{eq:thr} for $\ms$, the modulation necessary for the system to operate as a CPA or as a laser is given by
\begin{equation} \label{eq:msCPAmin}
    \msCPAmin \approx \frac{\rsgamma{\alpha}+\sqrt{\rsgamma{\alpha}^2+2\Delta_{\alpha}^2}}{\chiz I_{\alpha}\Omega/2}
\end{equation}
We highlight that Eq.~\eqref{eq:msCPAmin} can be evaluated solely based on static resonant states and thus allows us to estimate $\msCPAmin$ only based on parameters of the static scattering system and the known driving frequency $\Omega$. Furthermore, since we express the above equation in terms of the frequency and intensity of the static resonant states, Eq.~\eqref{eq:msCPAmin} is general in the sense that it not only holds for the slab but for arbitrarily shaped objects.

To check the validity of the approximations used in the resonant state expansion, we compare Eq.~\eqref{eq:msCPAmin} (purple line) with an eigenvalue analysis of the associated $\SF$ using the TBM approximation in Fig.~\ref{fig:slab_CPA_modulation} (green line). As expected, both approaches give nearly identical results. For reference, we also show the results of a full-scale simulation (black line), in which we track the eigenvalues of $\SF$ using $n \in [-8,7]$ Floquet channels. In general, we observe that Eq.~\eqref{eq:msCPAmin} gives a reasonable approximation for $\msCPAmin$, allowing us to adequately estimate the CPA-lasing modulation strength solely based on static information. More specifically, we observe that all three lines (RSE, TBM, full simulation) show a similar behavior: For those normalized thicknesses, where only a weak modulation strength $\depsEPmin$ is necessary for the system to form an EP (blue line, same as in Fig.~\ref{fig:slab_EP_modulation}), also the modulation strength for the system to act as a CPA or a laser $\msCPAmin$ is reduced. Conversely, when a strong modulation is already necessary to reach an EP, a large amplitude is also needed for the system to operate as a CPA and laser.

\subsection{Data Availability}
The data that support the plots within this paper are available from the corresponding author on reasonable request.

\begin{acknowledgments}
The authors acknowledge helpful discussions with V.~Flynn, J.~Gohsrich, P.~A.~Huidobro, L.~Rebholz, and M.~Verde. D.G., J.H., and S.R.~acknowledge support from the Austrian Science Fund (FWF) under project P32300 (WAVELAND). A.C.V.~acknowledges support by the project No. 1.1.1.9/LZP/1/24/101: ``Non-Hermitian physics of spatiotemporal photonic crystals of arbitrary shape (PROTOTYPE)'' and the Visiting Awards for High Potentials from the University of Graz. P.G.~and C.R.~are part of the Max Planck School of Photonics, supported by the Bundesministerium für Bildung und Forschung, the Max Planck Society, and the Fraunhofer Society. P.G.~acknowledges support from the Karlsruhe School of Optics and Photonics (KSOP). P.G.~and C.R.~acknowledge support by the German Research Foundation within the SFB 1173 (project ID no. 258734477). 
\end{acknowledgments}

\setcounter{section}{0}
\renewcommand{\thesection}{S\arabic{section}}
\setcounter{figure}{0}
\renewcommand{\thefigure}{S\arabic{figure}}
\setcounter{equation}{0}
\renewcommand{\theequation}{S\arabic{equation}}

\newpage
\onecolumngrid
\begin{center}
    \LARGE \textbf{Supplementary Information}
\end{center}
\vspace{1em}
\twocolumngrid

\section{Connection of $\SF$ to Band Structure}
\label{app:bandstructures}

We mention here that the Floquet scattering matrix also provides access to the band structure of the underlying scattering systems \cite{Wang25}, which we interpret here as the spectral location their resonant states in the parameter space. The resonant states of a photonic structure are the self-standing modes that exist without any incident excitation. In general, for planar geometries, the following relation holds

\begin{eqnarray}
{S}^{-1}_\mathrm{F}(\omega, \mathbf{k}_\parallel)\ket{\cout}= \ket{\cin}\,  \label{eq:Smatplanar}
\end{eqnarray}

For such a resonant state, we require $\ket{\cout}\neq0$ given $\ket{\cin}=0$. Such a condition holds if

\begin{eqnarray}
\sigma_\mathrm{min}\{{S}^{-1}_\mathrm{F}(\omega, \mathbf{k}_\parallel)\}\rightarrow0\, \label{eq:bandstr}
\end{eqnarray}

where $\sigma_\mathrm{min}$ represents the minimum singular value. Those values of $\omega$ and $\mathbf{k}_\parallel$ that satisfy Eq.~\eqref{eq:bandstr} constitute the location of an eigenmode in the parameter space. Note that $\omega$ is in general complex due to the radiation and material losses of finite scattering structures.

\section{Further Details on the CPA and Lasing States \label{app:photonconserv}}

\begin{figure}
\includegraphics[width= 0.99\columnwidth]{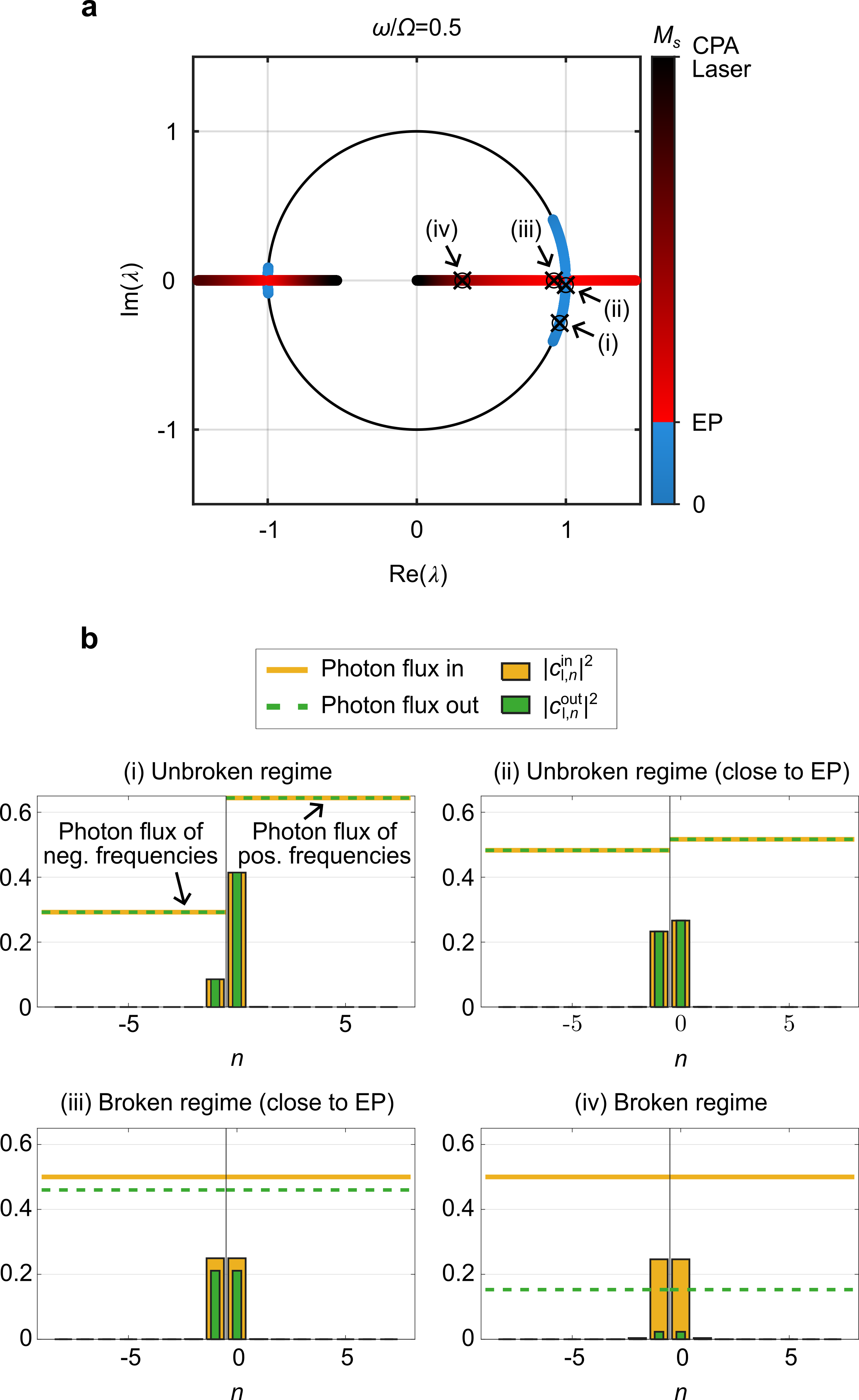}
\caption{\textbf{a} We track four eigenvalues of $\SF$ as a function of the modulation strength. For small $\ms$ all eigenvalues are unimodular and thus lie on the unit circle (blue). At a critical modulation strength, EPs are formed, where two respective eigenvalues coincide. By further increasing $\ms$, these two eigenvalues leave the unit circle in a pairwise fashion (red). \textbf{b}~We select four eigenvalues indicated by black crosses in \textbf{a} and plot the distribution of the components of the corresponding eigenstates of $\SF$ (orange bars). Furthermore, we depict the distribution of the associated output light field (green bars). The orange solid line represents the respective photon flux of the incoming state and the green dashed line represents the photon flux of the outgoing state. In panels (i) and (ii) we choose states from the unbroken symmetry regime, while panels (iii) and (iv) show states from the broken regime.}
\label{Fig:CPALasing_states}
\end{figure}

Here, we provide further details on the spatial and spectral distribution of the eigenstates of the Floquet scattering matrix for the periodically time-varying slab at normal incidence. Furthermore, we analyze the property of pseudophoton and photon conservation for eigenstates of $\SF$.

In Fig.~\ref{Fig:CPALasing_states}a, we depict the evolution of the four eigenvalues of the Floquet scattering matrix expressed in a photon-flux normalized basis that undergo a symmetry breaking transition. It contains the same data as Fig.~1d of the main text. Here, we also choose the same parameters for the slab as in the main text. We select four eigenvalues at different modulation amplitudes indicated by the black crosses (two in the unbroken regime and two in the broken regime) and investigate the associated eigenstates in detail.

In Fig.~\ref{Fig:CPALasing_states}b, we depict the corresponding eigenstates to these four eigenvalues. In panel~(i), we plot the distribution of the entries of the photon-flux normalized eigenstate corresponding to a unimodular eigenvalue and the associated outgoing state. Specifically, we plot the entries corresponding to the left asymptotic region, i.e., $\abs*{\cin_\indslabL}^2$ (orange) and $\abs*{\cout_\indslabL}^2$ (green). Due to the spatial symmetries of the scattering system, the distribution of the entries of the eigenstate for the right asymptotic region is identical to the left one. Therefore, we only plot the distribution corresponding to the left asymptotic region. In panel~(i), we observe that the distributions of the incoming and outgoing coefficients are the same, as expected for an eigenstate to a unimodular eigenvalue $\lambda_\nuz$ of the Floquet scattering matrix:
\begin{subequations}
\begin{align}
        \coutv_\nuz &= \SF \cinv_\nuz = \lambda_\nuz \cinv_\nuz \\
        \Rightarrow \abs{\left(\coutv_\nuz\right)_n}^2 &= \abs{\lambda_\nuz}^2 \abs{\left(\cinv_\nuz\right)_n}^2 = \abs{\left(\cinv_\nuz\right)_n}^2  \,  \label{eq:photonflux_unbroken}
\end{align}
\end{subequations}

Furthermore, we see that the eigenvector consists of both positive- and negative-frequency components. Interestingly, for the specific state depicted in panel~(i) of Fig.~\ref{Fig:CPALasing_states}b the corresponding incoming photon flux associated with the positive-frequency channels given by $\sum_{n \geq 0} \abs*{\left(\cinv_\nuz\right)_n}^2$ is larger than the corresponding negative-frequency photon flux $\sum_{n < 0} \abs*{\left(\coutv_\nuz\right)_n}^2$ (orange solid lines). In accordance with Eq.~\eqref{eq:photonflux_unbroken}, however, we see that the outgoing photon flux (green dashed lines) is the same as the incoming one. Thus, each eigenstate in the unbroken regime individually conserves the number of photons during scattering. Furthermore, the incoming and outgoing photon fluxes associated with the positive and negative frequencies are the same, confirming that the number of pseudophotons is conserved for this state, such that
\begin{equation}
    \sum_{n} \sign(\omega_n) \abs{\left(\coutv_\nuz\right)_n}^2 = \sum_{n} \sign(\omega_n) \abs{\left(\cinv_\nuz\right)_n}^2 \, 
\end{equation}

In panel~(ii), we depict an eigenstate corresponding to an eigenvalue that is close to the EP but still in the unbroken regime. We again observe that the incoming and outgoing photon flux is the same, indicating that this state conserves the number of photons (and also the number of pseudophotons). We highlight that being close to the EP, this state has nearly balanced photon flux associated with the positive- and negative-frequency components.

In panel~(iii), we show an eigenstate in the broken regime still being close to the EP. We see that the distribution of the components of the outgoing state is a rescaled (by a factor $\abs{\lambda_\num}<1$) version of the distribution of the incoming eigenstate. We can understand this by again considering the eigenvalue relation:
\begin{subequations}
\begin{align}
        \coutv_\num &= \SF \cinv_\num = \lambda_\num \cinv_\num \\
        \Rightarrow \abs{\left(\coutv_\nuz\right)_n}^2 &= \abs{\lambda_\num}^2 \abs{\left(\cinv_\nuz\right)_n}^2 \, \label{eq:photonflux_unbroken2}
\end{align}
\end{subequations}
Consequently, we observe that the number of photons is not conserved but rather reduced as the green dashed lines lies below the orange solid line. Importantly, however, we observe that the photon flux associated with the negative-frequency channels is identical to the photon flux of the positive-frequency channels. This shows that the number of pseudophotons is zero for both the incoming and the outgoing field and thus conserved, as we expect for an eigenstate of $\SF$ in the broken regime.

In panel~(iv), we demonstrate that as the absolute value of the eigenvalue $\abs{\lambda_\num}$ becomes smaller (approaching zero), the corresponding number of photons of the output light field decreases. At the CPA operational condition where $\abs{\lambda_\num} = 0$ (not explicitly shown here), no output light field is produced such that the outgoing photon flux vanishes. Conversely, the number of photons of an eigenvalue $\abs{\lambda_\nup}$ is increased during scattering with the extremal case of $\abs{\lambda_\nup} = \infty$ corresponding to the scenario of extreme photon amplification of the light field. Finally, also for the state depicted in panel~(iv), the photon flux is equally distributed among positive and negative channels. This indicates that this eigenstate has a vanishing pseudophoton content, as is characteristic of eigenstates of $\SF$ in the broken symmetry regime.

Finally, we want to clarify the implications of the above discussion for the energy flux of the associated states. A photon-flux normalized state $\cinoutv$ can be transformed into an energy-normalized state  $\dinoutv$ via the matrix $B$ as
\begin{equation}
    \dinoutv =B \cinoutv \, 
\end{equation}
where for the time-varying slab $B$ takes the form
\begin{equation}
\begin{split}
    B=\mathrm{diag}(&\dots,\sqrt{\hbar \abs{\omega_{-1}}},\sqrt{\hbar \abs{\omega_{0}}},\dots, \\
    &\dots, \sqrt{\hbar \abs{\omega_{-1}}},\sqrt{\hbar \abs{\omega_{0}}},\dots) \, 
\end{split}
\end{equation}
We use this connection for the following observation:
\begin{equation}
\begin{split}
    (\doutv_\nu)^\dagger \doutv_\nu &= (\coutv_\nu)^{\dagger}B^\dagger B \coutv_\nu \\
    &=\abs{\lambda_\nu}^2 (\cinv_\nu)^\dagger B^\dagger B\cinv_\nu = \abs{\lambda_\nu}^2 (\dinv_\nu)^\dagger \dinv_\nu \, 
\end{split}
\end{equation}
This relation tells us that whenever an eigenstate of $\SF$ conserves the number of photons (unbroken regime, $\nu=\nuz$), it also conserves energy, given by $\sum_n \abs*{\left(\dinv_\nu\right)_n}^2=\sum_n \abs*{\left(\doutv_\nu\right)_n}^2$. Conversely, a state from the broken regime ($\nu=\nupm$) neither conserves the number of photons nor its energy content.

Importantly, due the non-orthogonality of the eigenstates of the pseudounitary Floquet scattering matrix, even in the unbroken regime where all eigenstates individually conserve the number of photons and thus their energy content, energy (and the number of photons) is not a conserved quantity for arbitrary states. Superpositions of eigenstates from the unbroken regime will, in general, change their energy (number of photons) upon scattering. On the contrary, the number of pseudophotons is a conserved quantity of any superposition of eigenstates (irrespective if the constituents belong to the unbroken or broken regime).

\section{Far-field Intensity Calculations for the CPA and Lasing Points of the Time-varying Metasurface \label{app:MS_int}}
\begin{figure}[htbp]
\includegraphics[width= 0.99\columnwidth]{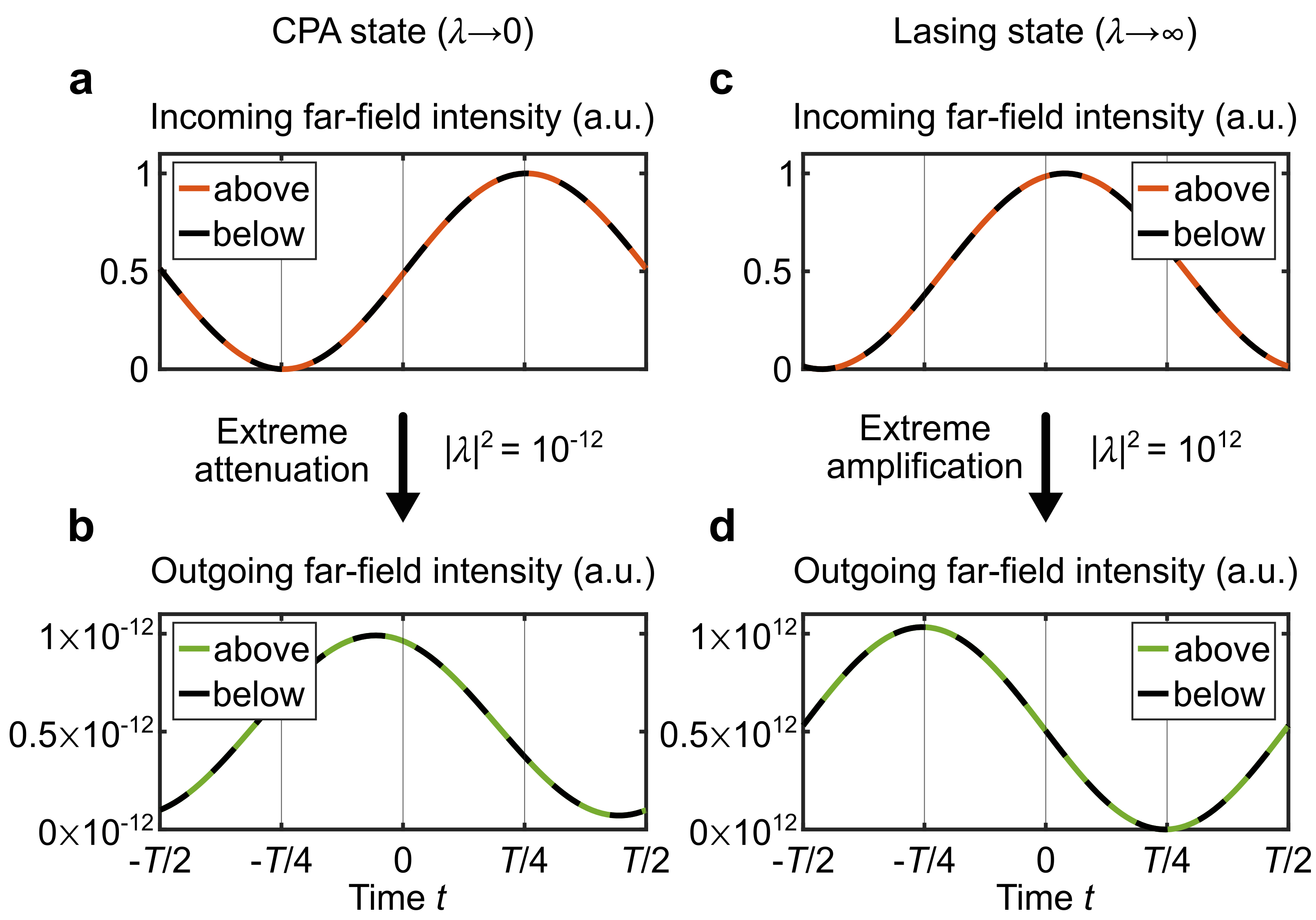}
\caption{Space-integrated incoming and outgoing intensities for the CPA and lasing states of the time-varying metasurface (see main text for parameter values) as a function of time $t$. We evaluate the electric fields in the far-field at $z=\pm1000 \times Tc$ away from the metasurface and spatially integrate them in the $x-y$ plane. \textbf{a}~Incoming intensity above the metasurface at $z=1000 \times Tc$ (red full line) and below the metasurface at $z=-1000 \times Tc$ (black dashed line) and \textbf{b}~outgoing intensity above (green full line) and below (black dashed line) corresponding the CPA state with an eigenvalue $\abs{\lambda}= 10^{-6}$. An extreme attenuation of the intensity can be observed, and nearly no outgoing field is produced. \textbf{c}~and~\textbf{d}~Same as~\textbf{a}, \textbf{b} but for the lasing state corresponding to an eigenvalue $\abs{\lambda}=10^{6}$. Here, we observe an extreme amplification of the incoming light field. We again see the time-reversal symmetry between the incoming CPA and outgoing lasing field.
}
\label{fig:intensity_MS}
\end{figure}

We plot the temporal intensity distribution in the far-field of the eigenstate corresponding to the minimal eigenvalue (CPA state) in Fig.~\ref{fig:intensity_MS}a-b. Specifically, we evaluate the electric field at $z=1000 \times Tc$ above (full lines) and below (dashed lines) the metasurface in the far-field and spatially integrate the intensity in this $x-y$ plane. The associated eigenvalue is $\abs{\lambda}=10^{-6}$, resulting in a drastic absorption of the input power such that nearly no outgoing field is produced. Conversely, in Fig.~\ref{fig:intensity_MS}c-d, we depict the far-field intensity of the eigenstate to the maximal eigenvalue (lasing state). There, we observe a large amplification of the incoming field corresponding to the eigenvalue $\abs{\lambda}=10^{6}$ of $\SF$. Again, the time-reversal symmetry between the CPA and the lasing state is apparent (the incoming CPA field is the time-reversed of the outgoing lasing field, and the incoming lasing field is the time-reversed of the outgoing CPA field).

\section{Floquet Scattering Matrix of the Time-varying Sphere \label{app:sphere}}
We use the Mie theory-based T-matrix method to evaluate $\SF$ of the time-varying sphere. We expand the incident field ${\mathbf{E}}^{\mathrm{inc}}(\mathbf{r}, t)$, internal field ${\mathbf{E}}^{\mathrm{int}}(\mathbf{r}, t)$,  and scattered field ${\mathbf{E}}^{\mathrm{sca}}(\mathbf{r}, t)$ for the time-varying sphere in a basis of vector spherical waves (VSWs) as
\begin{subequations}\label{eq:inc_sca}
\begin{align}
    {\mathbf{E}}^{\mathrm{inc}}(\mathbf{r},t)&=\sum_{nlm s}\tilde{a}_{nlm s}^{\mathrm{inc}}\mathbf{F}^{(1)}_{lms}(k_n\mathbf{r})\mathrm{e}^{-i\omega_n t}\,\label{eq:E_inc}\\
    {{\mathbf{E}}^{\mathrm{int}}(\mathbf{r},t)}&{=\sum_{nlm s}\left[\sum_{i}\tilde{a}_{ilm s}^{\mathrm{int}}S_{\kappa_i}(\omega_n)\mathbf{F}^{(1)}_{lms}(\kappa_i\mathbf{r})\right]\mathrm{e}^{-i\omega_n t}\,}\label{eq:E_int}\\
    {\mathbf{E}}^{\mathrm{sca}}(\mathbf{r},t)&=\sum_{nlm s}\tilde{a}_{nlm s}^{\mathrm{sca}}\mathbf{F}^{(3)}_{lm s}(k_n\mathbf{r})\mathrm{e}^{-i\omega_nt}\,\label{eq:E_sca}
    \end{align}
    \end{subequations}
    \noindent
where $k_n= \omega_n/c$. Here, $\tilde{a}_{nlm s}^{\mathrm{inc}},\tilde{a}_{ilm s}^{\mathrm{int}},\textrm{ and } \tilde{a}_{nlm s}^{\mathrm{sca}}$ represent the incident, internal, and scattered field coefficients, respectively. Furthermore, $\mathbf{F}^{(1)}_{lm s}(k_n\mathbf{r})$ [$\mathbf{F}^{(3)}_{lm s}(k_n\mathbf{r})$] represent the regular (radiating) VSWs with total angular momentum $l=1,2,3...,l_\mathrm{max}$; $z$-component of angular momentum $m= -l, -l+1,...,l$; and parity $s= 0,1$. Here, $s=0$ represents the transverse-electric (TE), and $s=1$ represents the transverse-magnetic (TM) mode, respectively. Moreover, $l_\mathrm{max}$ is the maximum multipolar order used in the expansion. Furthermore, $\kappa^2_i$ and $\ket{S_{\kappa_i}}$ represent the eigenvalues and the corresponding eigenvectors, respectively, of the frequency coupling matrix of a time-varying homogeneous medium. The explicit expression of the used frequency coupling matrix is given by \cite[Eq.~(10)]{ptitcyn2023floquet}. Here, $i\in[-N-1,N]$. Note that $\ket{S_{\kappa_i}}$ has $2(N+1)$ entries and $S_{\kappa_i}(\omega_j)$ corresponds to its $j^\textrm{th}$ entry.

Next, to find the relation between $\ket*{\tilde{a}^\mathrm{inc}}$ and $\ket*{\tilde{a}^\mathrm{sca}}$, we use the following interface conditions on the surface of the sphere \cite{ptitcyn2023floquet}

\begin{subequations}
\begin{eqnarray}
\hspace{-9mm}
{
\left.\hat{\mathbf{r}}\!\times\!
\left[{\mathbf{E}}^{\mathrm{int}}(\mathbf{r},t)
-{\mathbf{E}}^{\mathrm{sca}}(\mathbf{r},t)
-{\mathbf{E}}^{\mathrm{inc}}(\mathbf{r},t)\right]
\right|_{\mathrm{r}=R}
}
&{=}& {0\,} \label{eq:BC1}\\[2mm]
\hspace{-9mm}
{
\left.\hat{\mathbf{r}}\!\times\!
\left[{\mathbf{H}}^{\mathrm{int}}(\mathbf{r},t)
-{\mathbf{H}}^{\mathrm{sca}}(\mathbf{r},t)
-{\mathbf{H}}^{\mathrm{inc}}(\mathbf{r},t)\right]
\right|_{\mathrm{r}=R}
}
&{=}& {0\,} \label{eq:BC2}
\end{eqnarray}
\end{subequations}

Here, ${\mathbf{H}}^{\mathrm{inc}}(\mathbf{r},t)$, ${\mathbf{H}}^{\mathrm{int}}(\mathbf{r},t)$, and ${\mathbf{H}}^{\mathrm{sca}}(\mathbf{r},t)$ represent the incident, internal, and scattered magnetic field, respectively. These interface conditions allow us to link the coefficients $\ket*{\tilde{a}^\mathrm{inc}}$ and $\ket*{\tilde{a}^\mathrm{sca}}$ by the Floquet T-matrix $\tilde{T}_0$ as

\begin{eqnarray}
\ket*{\tilde{a}^\mathrm{sca}}= \tilde{T}_0\ket*{\tilde{a}^\mathrm{inc}}\, .\label{eq:Tmat}
\end{eqnarray}
Note that $\tilde{T}_0$ is a square matrix with the dimension $2l_\mathrm{max}(2N)(l_\mathrm{max}+2)$. 

Here, for a given $\ket*{\tilde{a}^\mathrm{sca}}$, the scattered power ${P}_n^{\mathrm{sca}}$ for the frequency $\omega_n$ in the far-field is given by \cite{ptitcyn2023floquet}
\begin{eqnarray}
{P}_n^{\mathrm{sca}}=\sum_{lms}\frac{c^2\left|{\tilde{a}}^{\mathrm{sca}}_{nlms}\right|^2}{{Z}_0\omega_n^2}\,
\label{eq:Psca}
\end{eqnarray}
Therefore, the photon flux $\phi_n$ corresponding to ${P}_n^{\mathrm{sca}}$ reads
\begin{eqnarray}
\phi_n=\frac{{P}_n^{\mathrm{sca}}}{\hbar |\omega_n|}=\sum_{lms}\frac{c_0^2\left|{\tilde{a}}^{\mathrm{sca}}_{nlms}\right|^2}{{Z}_0\hbar|\omega_n|^3}\,
\label{eq:fluxsph}
\end{eqnarray}
where $\hbar$ is the reduced Planck's constant, and $Z_0$ is the impedance of free space. From Eq.~\eqref{eq:fluxsph}, we note that the scattered field coefficients $\ket*{{a}^\mathrm{sca}}$ in a photon-flux normalized basis can be obtained by the transformation

\begin{eqnarray}
{{a}}^{\mathrm{sca}}_{nlms}=\sqrt\frac{c^2}{Z_0\hbar|\omega_n|^{3}}{\tilde{a}}^{\mathrm{sca}}_{nlms}\,
\label{eq:asc_sph}
\end{eqnarray}
Similarly, for the incident field coefficients $\ket*{{a}^\mathrm{inc}}$, we can write 
\begin{eqnarray}
{{a}}^{\mathrm{inc}}_{n'l'm's'}=\sqrt\frac{c^2}{Z_0\hbar|\omega_{n'}|^{3}}\tilde{a}^{\mathrm{inc}}_{n'l'm's'}\,
\label{eq:ain_sph}
\end{eqnarray}
Furthermore, combining Eqs.~\eqref{eq:Tmat}, \eqref{eq:asc_sph}, and \eqref{eq:ain_sph}, we get
\begin{eqnarray}
\ket*{{a}^\mathrm{sca}}=T_0\ket*{{a}^\mathrm{inc}}\, .\label{eq:Tmatnorm}
\end{eqnarray}
Here, $T_0$ is the Floquet T-matrix of the sphere in a photon-flux normalized basis. The elements of $T_0$ can be calculated from those of $\tilde{T}_0$ by using the transformation \cite{Lamprianidis}
\begin{eqnarray}
{T_0}^{\{nlms\},\{n'l'm's'\}}=\sqrt{\left| \frac{\omega_{n'}}{\omega_n} \right|^3}{\tilde{T}_0}^{\{nlms\},\{n'l'm's'\}}.
\end{eqnarray}
Here, the first set of indices enumerates rows, and the second set enumerates columns of the Floquet T-matrices. 

Finally, we can compute the Floquet scattering matrix of the time-varying sphere in a photon-flux normalized basis as \cite{waterman2009tmatrix}
\begin{equation}
\label{eq:Smat_sph}
S_\mathrm{F}=\mathbb{1}+2T_0\,
\end{equation}
\noindent
Note that the Floquet scattering matrix $S_\mathrm{F}$ satisfies Eq.~\eqref{eq:pseudounitarity}. Furthermore, $S_\mathrm{F}$ also satisfies Eq.~\eqref{eq: Smat_cc_sym} for permittivity profiles $\epsilon(t)$ that are symmetric with respect to $t=0$.

\section{Floquet Scattering Matrix of the Time-varying Metasurface \label{app:metasurface}}
We expand the incoming field ${\mathbf{E}}^{\mathrm{in}}(\mathbf{r}, t)$ and the outgoing field ${\mathbf{E}}^{\mathrm{out}}(\mathbf{r}, t)$ from the time-varying metasurface for the Floquet frequency $\omega$ and Bloch wavevector $\mathbf{k}_\parallel$ in a basis of plane waves as \cite{garg2022modeling,beutel2021efficient}
\begin{subequations}
\begin{align}
    \mathbf{E}^{\mathrm{in}}(\mathbf{r}, t) &= \sum_{n\mathbf{g} \alpha d} \tilde{u}^{\mathrm{in}}_{n\mathbf{g} \alpha d} \, \mathbf{P}_{\mathbf{g} \alpha d}(k_n \mathbf{r}) \, \mathrm{e}^{-i \omega_n t} \, \\
    \mathbf{E}^{\mathrm{out}}(\mathbf{r}, t) &= \sum_{n\mathbf{g} \alpha d}\tilde{u}^{\mathrm{out}}_{n\mathbf{g} \alpha d} \, \mathbf{P}_{\mathbf{g} \alpha d}(k_n \mathbf{r}) \, \mathrm{e}^{-i \omega_n t} \,
\end{align}
\end{subequations}
where $\mathbf{P}_{\mathbf{g} \alpha d}(k_n\mathbf{r})$ corresponds to a plane wave with the wavevector $\mathbf{k}_{n\mathbf{g}\alpha d}=(\mathbf{k}_\parallel+\mathbf{g})+(-1)^{d}\sqrt{k_n^2-(\mathbf{k}_\parallel+\mathbf{g})^2}\,\hat{\mathbf{z}}$, and polarization $\alpha$. Here, $\alpha$ takes the values $0$ and $1$ for TE and TM polarized plane waves, respectively. Furthermore, $d$ takes the values $0$ and $1$ for downward and upward propagating plane waves (with respect to the plane of the metasurface), respectively. Moreover, $\mathbf{g}$ corresponds to a reciprocal lattice vector. In the expansion of the fields $\mathbf{E}^{\mathrm{in}}(\mathbf{r}, t)$ and $\mathbf{E}^{\mathrm{out}}(\mathbf{r}, t)$, we assume $G_n$ as the total number of diffraction orders for the frequency $\omega_n$. Here, $G_n$ is chosen such that $|k_n|>|(\mathbf{k}_\parallel+\mathbf{g})|$ for all $\omega_n$. This choice ensures that only the contribution of propagating plane waves is taken into account when evaluating $\mathbf{E}^{\mathrm{in}}(\mathbf{r}, t)$ and $\mathbf{E}^{\mathrm{out}}(\mathbf{r}, t)$. Such a choice of $G_n$ is justified as it implies that the sources and the detectors for $\mathbf{E}^{\mathrm{in}}(\mathbf{r}, t)$ and $\mathbf{E}^{\mathrm{out}}(\mathbf{r}, t)$ are placed in the far-field of the metasurface. 

We connect the plane wave coefficients $\ket*{\tilde{u}^\mathrm{in}}$ and $\ket*{\tilde{u}^\mathrm{out}}$ using the Floquet scattering matrix $\tilde{S}_\mathrm{F}$ of the metasurface as \cite{garg2022modeling}

\begin{eqnarray}
\ket*{\tilde{u}^\mathrm{out}}= \tilde{S}_\mathrm{F}\ket*{\tilde{u}^\mathrm{in}}\, . \label{eq:Smatmeta}
\end{eqnarray}
Here, $\tilde{S}_\mathrm{F}$ is a square matrix with dimension $\sum_{n=-N}^{N-1}4G_n$.

Note that we numerically evaluate $\tilde{S}_\mathrm{F}$ using the effective Floquet T-matrix $\tilde{T}_\mathrm{eff}$ of the metasurface. Using the Floquet T-matrix of an isolated time-varying sphere, $\tilde{T}_\mathrm{eff}$ can be calculated as

\begin{equation}
{\tilde{T}_\mathrm{eff}(\omega, \mathbf{k_{\parallel}})= \left[\mathbb{1}-{\tilde{T}_{0}(\omega)}\sum_{\mathbf{R}\neq0}\tilde{C}^{(3)}(-\mathbf{R})\hspace{1pt}\mathrm{e}^{i\mathbf{k_{{\parallel}}}\cdot\mathbf{R}}\right]^{-1}{\tilde{T}_{0}(\omega)}\,}\label{eq:Teff}
\end{equation}

Here, $\tilde{C}^{(3)}$ is the translation matrix of VSWs for time-varying metasurfaces (see \cite[Eq.~(S16)] {garg2022modeling}).

To evaluate $\tilde{S}_\mathrm{F}$, from $\tilde{T}_\mathrm{eff}$, we use 

\begin{equation}
\label{eq:Smat_meta_tmat}
{\tilde{S}_\mathrm{F}=\mathbb{1}+\tilde{O}\hspace{2pt}{\tilde{T}_\mathrm{eff}}\hspace{2pt}\tilde{I}\,}
\end{equation}
\noindent
Here, $\tilde{{O}}$ is the coordinate transformation matrix of the scattered field from the spherical wave to the plane wave basis, and $\tilde{{I}}$ is the coordinate transformation matrix of the incident field from the plane wave to the spherical wave basis \cite[Eqs.~(S8), (S13)]{garg2022modeling}.

Next, we write the power flux $P_n^\mathrm{out}$ carried by the outgoing fields $\mathbf{E}^{\mathrm{out}}(\mathbf{r}, t)$ for the frequency $\omega_n$ in the far-field as \cite{garg2022modeling}
\begin{equation}
      P^n_\mathrm{out}=\frac{c}{2Z_0}\sum_{\mathbf{g}\alpha d}\frac{|k_{z,n\mathbf{g}\alpha d}|}{|\omega_n|}|\tilde{u}^\mathrm{out}_{n\mathbf{g}\alpha d}|^2\,
\end{equation}
Here, $k_{z,n\mathbf{g}\alpha d}$ refers to the $z$-component of the wavevector $\mathbf{k}_{n\mathbf{g}\alpha d}$. The photon flux $\phi_n$ corresponding to $P_n^\mathrm{out}$ is given by 

\begin{eqnarray}
\phi_n=\frac{c}{2Z_0}\sum_{\mathbf{g}\alpha d}\frac{|k_{z,n\mathbf{g}\alpha d}|}{\hbar \omega_n^2}|\tilde{u}^\mathrm{out}_{n\mathbf{g}\alpha d}|^2\,
\label{eq:fluxmeta}
\end{eqnarray}
From Eq.~\eqref{eq:fluxmeta}, we note that the outgoing field coefficient $\ket*{u^\mathrm{out}}$ in a photon-flux normalized basis can be obtained by the transformation

\begin{eqnarray}
{{u}}^{\mathrm{out}}_{n\mathbf{g}\alpha d}=\sqrt\frac{c|k_{z,n\mathbf{g}\alpha d}|}{2Z_0\hbar\omega_n^{2}}{\tilde{u}}^{\mathrm{out}}_{n\mathbf{g}\alpha d}\,
\label{eq:aout_meta}
\end{eqnarray}
Similarly, for the incoming field coefficients $\ket*{u^\mathrm{in}}$, we can write 
\begin{eqnarray}
{{u}}^{\mathrm{in}}_{n'\mathbf{g}'\alpha' d'}=\sqrt\frac{c|k_{z,n'\mathbf{g}'\alpha' d'}|}{2Z_0\hbar(\omega_n')^{2}}{\tilde{u}}^{\mathrm{in}}_{n'\mathbf{g}'\alpha' d'}\, \label{eq:ain_meta}
\end{eqnarray}
Furthermore, combining Eqs.~\eqref{eq:Smatmeta}, \eqref{eq:aout_meta}, and \eqref{eq:ain_meta}, we get
\begin{eqnarray}
\ket*{u^\mathrm{out}}= S_\mathrm{F}\ket*{u^\mathrm{in}}\, .\label{eq:Smatnorm_meta}
\end{eqnarray}
Here, $S_\mathrm{F}$ is the Floquet scattering matrix of the metasurface in a photon flux normalized basis. The elements of $S_\mathrm{F}$ can be calculated from those of $\tilde{S}_\mathrm{F}$ using the transformation
\begin{equation}
\begin{split}
{S}_\mathrm{F}^{\{n\mathbf{g}\alpha d\},\{n'\mathbf{g}'\alpha'd'\}} =&
\left| \frac{\omega_{n'}}{\omega_n} \right|
\sqrt{ \left|\frac{k_{z,n\mathbf{g}\alpha d}}{k_{z,n'\mathbf{g}'\alpha'd'}} \right|} \\
&\times\tilde{S}_\mathrm{F}^{\{n\mathbf{g}\alpha d\},\{n'\mathbf{g}'\alpha'd'\}}\, 
\end{split}
\end{equation}
We arrange $\ket{u^\mathrm{out}}$ and $\ket{u^\mathrm{in}}$ following the conventions: $\ket{u^\mathrm{out}}=(\textbf{u}^\mathrm{out}_\downarrow,\textbf{u}^\mathrm{out}_\uparrow)^\mathrm{T}$ and $\ket{u^\mathrm{in}}=(\textbf{u}^\mathrm{in}_\uparrow,\textbf{u}^\mathrm{in}_\downarrow)^\mathrm{T}$, respectively. Here, $\textbf{u}^\mathrm{out}_\uparrow$ ($\textbf{u}^\mathrm{in}_\uparrow$) correspond to the upward propagating outgoing (incoming) and $\textbf{u}^\mathrm{out}_\downarrow$ ($\textbf{u}^\mathrm{in}_\downarrow$) correspond to the downward propagating outgoing (incoming) field coefficients. Importantly, we arrange the entries of the $\SF$ matrix as,
\begin{equation}
    \SF=\mqty(S_{\downarrow\uparrow} & S_{\downarrow\downarrow} \\ S_{\uparrow\uparrow} & S_{\uparrow\downarrow})\, 
\end{equation} 
which is crucial for correctly describing the spectral locations of the EPs as shown in \cite{novitsky2020unambiguous}.
Note that $S_\mathrm{F}$ satisfies Eq.~\eqref{eq:pseudounitarity}. Furthermore, $S_\mathrm{F}$ also satisfies Eq.~\eqref{eq: Smat_cc_sym} for permittivity profiles $\epsilon(t)$ that are symmetric with respect to $t=0$.

\bibliography{literature} 

\end{document}